\documentclass[journal]{IEEEtran}

\IEEEoverridecommandlockouts

\usepackage[cmex10]{amsmath}
\usepackage{amssymb}
\usepackage{mdframed}
\usepackage{amsthm}
\usepackage{cases}
\usepackage{multirow}
\usepackage{empheq}

\usepackage{siunitx}

\usepackage[noadjust]{cite}
\usepackage{verbatim}

\usepackage{algorithm}
\usepackage{algpseudocode}

\ifCLASSINFOpdf
\usepackage[pdftex]{graphicx}
\else
\usepackage[dvips]{graphicx}
\fi
\usepackage[caption=false,font=footnotesize]{subfig}

\usepackage{color}
\usepackage{multirow}

\usepackage[hyphens]{url}

\begin{document}
	
    \sloppy
	
    \title{
    Token-Oriented Semantic Communication with Pretrained Vision Transformers
    }  
	
    \author{
		\IEEEauthorblockN{Jiwoong~Im, Minwoo~Kim, Jaeho~Lee,~\IEEEmembership{Member,~IEEE}, Yo-Seb~Jeon,~\IEEEmembership{Member,~IEEE}, and Yongjune~Kim,~\IEEEmembership{Member,~IEEE}}	\\		
    
    \thanks{
                
        Jiwoong Im, Minwoo Kim, Jaeho Lee, Yo-Seb Jeon, and Yongjune Kim are with the Department of Electrical Engineering, Pohang University of Science and Technology (POSTECH), Pohang 37673, South Korea (e-mail: \{jw3562, minwoo.kim, jaeho.lee, yoseb.jeon, yongjune\}@postech.ac.kr).
        Jiwoong Im and Minwoo Kim contributed equally to this work.
    }  

    }
	
    \maketitle
    	
    \begin{abstract}

    Token communications realize the semantic communication principle at the granularity of transformer tokens, providing a promising direction for client--server collaborative inference in resource-constrained edge systems.
    However, directly transmitting token embeddings presents two practical challenges: substantial communication cost and limited interoperability across model-specific token embedding spaces.
    To address these challenges, we propose a \emph{token-oriented} semantic communication framework. 
    In this framework, token-level task relevance determines which compressed image latents are transmitted, enabling token-granular transmission without directly transmitting token embeddings.    
    The framework is modular, coordinating three pretrained components---a lightweight client-side vision transformer (ViT), a learned image compression (LIC) model, and a large server-side ViT---without end-to-end training.
    The key enabler is the one-to-one spatial alignment between ViT patch tokens and the LIC latent vectors, which allows token-level task relevance to directly determine which latent vectors are transmitted.
    Building on this alignment, token-aligned LIC selectively transmits task-relevant latents, layer-selective attention rollout estimates token relevance from a selected range of attention layers in a single forward pass, and surrogate token substitution adapts the frozen server model by optimizing a single learnable token.
    Experiments on ImageNet show that the proposed framework achieves a more favorable rate--accuracy trade-off than recent semantic communication schemes, hand-crafted codecs, and task-agnostic LIC models.
    
    \end{abstract}  

	
    \section{Introduction}

    Recent advances in transformer architectures have fundamentally reshaped the internal representation and processing of modern artificial intelligence (AI) models~\cite{Vaswani2017attention}.
    Transformers process inputs as a sequence of tokens, each of which serves as a unified computational unit across multiple layers.
    This tokenized formulation has proven effective across a wide range of modalities, including language, images, audio, and video~\cite{Vaswani2017attention,Dosovitskiy2021image,Gong2021Ast,Arnab2021Vivit,Bertasius2021Space,Radford2021Learning}.
    In particular, token representations offer a structured, modular interface that is naturally compatible with global context modeling via attention mechanism~\cite{Vaswani2017attention}.
    As a result, tokens are now widely regarded not only as an internal feature format for transformer models, but also as a candidate unit for computation and communication.

    Meanwhile, with the emergence of AI-native wireless networks, communication paradigms are progressively shifting beyond conventional bit-level communication toward \emph{semantic communications}~\cite{Zhang2022goal,Shi2021from,Lan2021what,Gunduz2022beyond,Xie2020lite,Kim2023distributed,Kim2020distributed}.
    Rather than optimizing human-perceived reconstruction fidelity quantified in terms of bit-level distortion, semantic communications attempt to enhance communication efficiency by selectively transmitting information relevant to downstream task objectives.
    Tokens provide a natural unit for this shift, and recent studies have accordingly investigated \emph{token communications}, which realize the semantic communication principle at the granularity of transformer tokens~\cite{Qiao2025Token,Liu2025Text,Peng2025Large,Devoto2024Adaptive,Devoto2026Adaptive,Wei2026Token}.
    Operating at the token level reveals which tokens are relevant to the downstream task, so that the communication payload can be concentrated on inference-relevant tokens rather than the entire source, as in conventional bit-level communication.

    Realizing this advantage by transmitting token representations themselves, however, entails two practical considerations.
    The first concern is communication cost: each token consists of a high-dimensional embedding vector whose dimensionality increases with model capacity, so transmitting even a reduced number of tokens may incur a substantial payload.
    The second concern is the limited interoperability of token embedding spaces across independently trained models, as these embedding spaces can differ substantially because of differences in architectures, pretraining procedures, and downstream objectives, even at an identical model dimensionality~\cite{Kornblith2019Similarity}.
    To address both considerations, several studies jointly train the transmitter and receiver in an end-to-end manner, typically with a shared codebook that converts token embeddings into codeword indices~\cite{van-den-Oord2017Neural,Qiao2025Token,Hu2023Robust,Zhang2024Unified,Yu2024Image}.
    Such fully integrated designs are effective within a given deployment, yet they reduce modularity, incur retraining costs, and couple the communication strategy to a specific downstream task.

    To retain the benefits of token-level processing while addressing these considerations, we propose a \emph{token-oriented} semantic communication framework, considering image classification as the downstream task.
    Rather than compressing or exchanging token embeddings, the proposed framework compresses the input image with learned image compression (LIC), which exploits the statistics of natural images to attain a content-adaptive rate--distortion performance competitive with conventional image compression~\cite{Balle2017End,Balle2018Variational,Minnen2018Joint,He2022Elic}.
    The proposed framework integrates token-level task relevance into this compression.
    The key enabler is the spatial alignment between visual tokens and LIC latent vectors: in the widely used architectures, the vision transformer (ViT) patch size equals the LIC encoder’s spatial downsampling factor, so each visual token corresponds to exactly one latent vector~\cite{Dosovitskiy2021image,Touvron2021training,Touvron2022Deit,Balle2018Variational,Minnen2018Joint,He2022Elic}.
    Task relevance is thereby incorporated into compression without modifying the LIC architecture or introducing additional training, and the communication payload is reduced by compressing in the image domain rather than in the model-specific embedding space.
    We use the term \emph{token-oriented} to emphasize that token-level task relevance determines which token-aligned LIC latent vectors are transmitted, enabling transmission at token granularity without directly transmitting token embeddings.

    The resulting framework is fully \emph{modular}, coordinating three separately pretrained models for client--server collaborative inference: a lightweight client-side ViT, an LIC model, and a large server-side ViT.
    Specifically, the client-side ViT estimates the task relevance of visual tokens, and this token-level relevance is used to guide the selection of latent representations within the compressor.
    The token-aligned compressor then reduces communication payload by transmitting only the selected latent representations, while the server-side ViT performs downstream inference from the reconstructed image.
    Since no token embedding is exchanged across models, their embedding spaces need not be compatible, and each component can be replaced without retraining the others from scratch.

    Prior semantic communication frameworks that use token-level relevance to guide transmission without directly transmitting token embeddings can likewise be regarded as token-oriented~\cite{Liu2023efficient,Liu2024adaptive,Im2024Attention,Park2025Vision}.
    However, they differ in the transmitted representation, sending image patches in the pixel domain, either raw or scalar-quantized.
    Operating in the LIC latent domain thus distinguishes the proposed framework within this class and enhances the rate--accuracy trade-off.

    Upon this framework, we introduce three complementary components, each acting on one of the pretrained models, to further optimize the rate--accuracy trade-off:
    \begin{itemize}
        \item \emph{Token-aligned LIC}:
        Token-level task relevance is integrated into an LIC model to reduce the transmission rate while preserving task-relevant visual information.
        Exploiting the structural alignment between the visual tokens of ViT and the latent representations of LIC on a shared spatial grid, the token-level relevance map directly indexes the latents to be transmitted.
        Our approach selectively transmits only the latents at task-relevant positions, reducing the transmission rate while retaining compatibility with the underlying LIC architecture.
        
        \item \emph{Layer-selective attention rollout}:
        The client-side ViT estimates token-level task relevance precisely by aggregating attention weights over a selected range of transformer layers.
        This selective aggregation provides more accurate relevance estimates than the last-layer attention commonly adopted in prior semantic communication frameworks~\cite{Liu2023efficient,Liu2024adaptive,Im2024Attention,Park2025Vision}, the attention rollout accumulated over all layers~\cite{Abnar2020quantifying}, and even gradient-weighted attention~\cite{Chefer2021Generic}.
        Moreover, unlike gradient-weighted attention~\cite{Chefer2021Generic}, it requires only a single forward pass, avoiding the memory and latency overhead of gradient computation that is demanding for resource-constrained clients.
    
        \item \emph{Learnable surrogate token substitution}:
        Server-side inference with partial visual information is improved by substituting the token positions of the unselected regions with a single learnable \emph{surrogate token}.
        Inference without any adaptation, whether using the directly reconstructed image or only the selected patches, degrades classification accuracy on the server, whereas retraining the large server backbone is costly and alters its behavior on complete inputs.
        Instead, we optimize only the single surrogate token with the backbone kept frozen, providing parameter-efficient input-level adaptation that improves inference with partial visual information while preserving the original complete-input behavior.
        It further improves the robustness of server-side inference over unreliable channels, where the tokens affected by channel impairments are likewise replaced by the surrogate token.
    \end{itemize}    
    
    Built from three pretrained models without end-to-end training, the proposed framework enables modular and communication-efficient collaborative inference. 
    Across the evaluated rate regime, it attains a more favorable rate--accuracy trade-off than both recent semantic communication schemes~\cite{Im2024Attention,Park2025Vision}, which exploit task relevance without efficient compression, and task-agnostic LIC models~\cite{Balle2018Variational,Minnen2018Joint,He2022Elic}, which optimize reconstruction without task relevance.

    The remainder of this paper is organized as follows.
    Section~\ref{sec:related} reviews related work on token-level semantic communications, learned image compression, attention-guided importance estimations, and learnable input tokens.
    Section~\ref{sec:proposed} presents an overview of our proposed token-oriented semantic communication framework.
    Section~\ref{sec:main} introduces the main technical contributions of this work, optimizing the rate--accuracy trade-off of the proposed framework: token-aligned LIC, layer-selective attention rollout, and surrogate token substitution.
    Section~\ref{sec:experimental} reports the experimental results, and Section~\ref{sec:conclusion} concludes the paper.

    \section{Related Work}\label{sec:related}

    \subsection{Token-Level Semantic Communications}

    \begin{table*}[!t]
    \centering
    \small
    \renewcommand{\arraystretch}{1.2}
    \caption{Comparison of Token-Level Semantic Communication Frameworks for Collaborative Inference}
    \label{tab:framework-comparison}
    \begin{tabular}{|c|c|c|c|c|}
        \hline
        \multirow{2}{*}{Framework} & Transmitted & Token Importance & Server-Side & \multirow{2}{*}{Modularity} \\
         & Representation & Estimation & Adaptation & \\ \hline \hline
        VQ-Based Token Comm.~\cite{Qiao2025Token,Hu2023Robust,Zhang2024Unified,Yu2024Image} & VQ Codebook Indices & Auxiliary Network or None & Joint Training & $\times$ \\ \hline
        Prior Token-Oriented Comm.~\cite{Liu2023efficient,Liu2024adaptive,Im2024Attention,Park2025Vision} & Raw or Quantized Pixels & Last-Layer Attention & None & $\checkmark$ \\ \hline
        \multirow{2}{*}{\textbf{Proposed}} & \textbf{Token-Aligned} & \textbf{Layer-Selective} & \textbf{Surrogate Token} & \multirow{2}{*}{$\checkmark$} \\
         & \textbf{LIC Latents} & \textbf{Attention Rollout} & \textbf{(Frozen Backbone)} & \\ \hline
    \end{tabular}
    \end{table*}

    Semantic communication frameworks operating at token granularity differ in what they transmit, how they identify task-relevant content, and how the receiver is adapted for inference.
    As summarized in Table~\ref{tab:framework-comparison}, VQ-enabled token communications transmit the token representations themselves as vector-quantized codebook indices, which requires end-to-end joint training to establish a codebook shared between the two ends~\cite{Qiao2025Token,Hu2023Robust,Zhang2024Unified,Yu2024Image}.
    Token-oriented communications instead guide transmission by token-level relevance while operating on pretrained models, yet prior frameworks in this class transmit image patches in the pixel domain~\cite{Liu2023efficient,Liu2024adaptive,Im2024Attention,Park2025Vision}.
    The proposed framework is likewise token-oriented but transmits LIC latents, thereby improving rate efficiency while retaining modularity, i.e., operation with pretrained models without end-to-end joint training.

    \subsection{Learned Image Compression}

    With the advancement of machine learning, the learned image compression (LIC) paradigm~\cite{Balle2017End,Balle2018Variational,Minnen2018Joint,He2022Elic} has achieved rate--distortion performance competitive with conventional hand-crafted image compression.
    Among the various LIC architectures, we build upon the nonlinear transform coding framework~\cite{Balle2017End} and its extension to a hyperprior network~\cite{Balle2018Variational}, which underlies many widely adopted LIC models.
    The end-to-end compression pipeline maps an input image $\mathbf{X}$ to a reconstruction $\widehat{\mathbf{X}}\in \mathbb R^{H\times W\times C}$ through the following sequential pipeline:

    \begin{equation}\label{eq:LICpipeline}
        \mathbf{X} \xrightarrow{g_a} \mathbf{X}^c \xrightarrow{Q} \widehat{\mathbf{X}}^c \xrightarrow{g_s} \widehat{\mathbf X}.
    \end{equation}
    Here, an encoder $g_a$ extracts the latent representation $\mathbf{X}^c\in \mathbb R^{N\times D_c}$, where $D_c$ denotes the model dimension of the image encoder.
    $Q$ introduces discretization to obtain the quantized latent $\widehat{\mathbf{X}}^c$, and a decoder $g_s$ reconstructs the image.

    To effectively remove the remaining spatial redundancy of latents $\mathbf{X}^c$,~\cite{Balle2018Variational} introduced a hyperprior network that transmits side information.
    Specifically, a hyper encoder $h_a$ extracts a hyper latent $\mathbf{Z} = h_a(\mathbf{X}^c)$, which is quantized to $\widehat{\mathbf{Z}}$.
    A hyper decoder $h_s$ then decodes $\widehat{\mathbf{Z}}$ to estimate the conditional distribution of $\widehat{\mathbf{X}}^c$, modeled as a normal distribution with zero mean.
    Subsequently,~\cite{Minnen2018Joint} extended this architecture to jointly predict both the mean $\mu$ and variance $\sigma$ of the elements of the latents.
    These predicted parameters define the precise probability mass function required by the arithmetic coder, allowing the entropy coding process to closely approach the theoretical Shannon entropy limit.

    \subsection{Attention-Guided Importance Estimation}     

    Vision transformer (ViT) is a transformer-based architecture for computer vision tasks~\cite{Dosovitskiy2021image}.
    To perform inference, an input image $\mathbf{X} \in \mathbb{R}^{H\times W \times C}$ is reshaped into a sequence of flattened 2D patches $\mathbf{X}^p \in \mathbb{R}^{N \times (P^2\cdot C)} $, where $(H, W)$, $C$, and $(P, P)$ denote the resolution of the original image, the number of channels, and the resolution of each image patch, respectively. 
    Here, $N = \frac{HW}{P^2}$ denotes the resulting number of image patches.
    Each patch $\mathbf{x}_i^p$ is then tokenized into a visual token $\mathbf{x}_i^t \in \mathbb{R}^{1 \times D}$ by linear projection, forming the visual token matrix $\mathbf{X}^t \in \mathbb{R}^{N \times D}$ whose $i$th row is $\mathbf{x}_i^t$. 
    Then, the \emph{class token} $\mathbf{x}_\mathrm{cls}$ is prepended to the sequence of visual tokens.
    The input embedding of the ViT encoder $\widetilde{\mathbf{X}}^t \in \mathbb{R}^{(N+1) \times D}$ is given by
    \begin{equation}\label{eq:vit-input_sequence}
        \widetilde{\mathbf{X}}^t = \left[ \mathbf{x}_\mathrm{cls}; \mathbf{X}^t + \mathbf{E} \right],
    \end{equation}
    where $\mathbf{E} \in \mathbb{R}^{N \times D}$ denotes the standard learnable position embedding.

    Since token interactions in ViT are primarily mediated through multi-head self-attention blocks, the attention weights can be interpreted as modeling global dependencies among tokens that contribute to the final inference.
    Accordingly, many recent studies have estimated token relevance $\mathbf{R} \in \mathbb{R}^{(N+1) \times (N+1)}$ and token-level task relevance $\bar{\mathbf{r}} \in \mathbb{R}^{N}$ from the attention weights computed during downstream inference.
    Below, we briefly review representative attention-guided token relevance methods.
    For brevity, we denote the attention weights at layer $l$ and head $h$ by $\mathbf{A}_{l,h} \in \mathbb{R}^{(N+1) \times (N+1)}$.
    \begin{itemize}
        \item \emph{Last-layer attention}:
        The most direct approach estimates token relevance using the attention weights from the final transformer layer:
        \begin{equation}
            \mathbf{R} = \mathbb{E}_h \left[ \mathbf{A}_{L,h} \right],
        \end{equation}
        where $L$ denotes the number of transformer layers and $\mathbb{E}_h[\cdot]$ represents averaging over attention heads.
        It is widely used in prior token-oriented communication studies owing to its simplicity~\cite{Im2024Attention,Park2025Vision,Liu2023efficient,Liu2024adaptive}.

        \item \emph{Attention rollout}:
        Attention rollout aggregates attention weights across all transformer layers to approximate the overall token dependencies captured by the model~\cite{Abnar2020quantifying}:
        \begin{align}
            \widetilde{\mathbf{A}}_{l} = \mathbb{E}_h \left[ \mathbf{A}_{l,h} \right] + \mathbf{I}_{N+1}, \quad
            \mathbf{R} = \widetilde{\mathbf{A}}_{L} \widetilde{\mathbf{A}}_{L-1} \cdots \widetilde{\mathbf{A}}_{1},
        \end{align}
        where $\mathbf{I}_{N+1} \in \mathbb{R}^{(N+1) \times (N+1)}$ denotes the identity matrix that accounts for the residual connection in ViT.

        \item \emph{Gradient-weighted attention}:
        Gradient-weighted attention additionally reweights the attention rollout based on the gradient of the top-1 predicted logit with respect to the attention weights~\cite{Chefer2021Generic}.
        Although it provides precise task-relevance estimation, its reliance on gradient computation constitutes a practical limitation, which precludes its deployment on resource-constrained clients.
    \end{itemize}

    The above methods share a common final step: the token-level task relevance $\bar{\mathbf{r}}$ is extracted from the first row of $\mathbf{R}$, which quantifies the relevance of the class token, the input to the classification head, to the visual tokens.
    Excluding the class-token self-relevance $R_{1,1}$, we normalize the $N$ entries corresponding to the visual tokens as
    \begin{equation}\label{eq:relevance-normalization}
        \bar{r}_i = \frac{R_{1, i+1}}{\sum_{j=1}^N R_{1, j+1}},
    \end{equation}
    so that $\bar{\mathbf{r}}$ forms a distribution over the visual tokens.

    \subsection{Learnable Input Tokens}
    
    Learnable input tokens, i.e., trainable vectors inserted into the input sequence, have been used to adapt or train vision transformers.
    Visual prompt tuning prepends a few learnable \emph{prompt tokens} to the input and optimizes only them while freezing the backbone, enabling parameter-efficient task adaptation~\cite{Jia2022Visual}.
    On the other hand, masked image modeling inserts a single shared \emph{mask token} at masked positions, jointly optimized with the backbone for self-supervised reconstruction during pretraining~\cite{Bao2022Beit,He2022masked}.
    The proposed surrogate token substitution combines both aspects: a single learnable token substitutes missing positions and is optimized with the backbone frozen.
    It differs in purpose, being trained for the downstream classification objective to compensate for missing visual content at inference.

    \section{Overview of the Proposed Framework}\label{sec:proposed} 

    \begin{figure}[t] 
        \centering
        \includegraphics[width=0.42\textwidth]{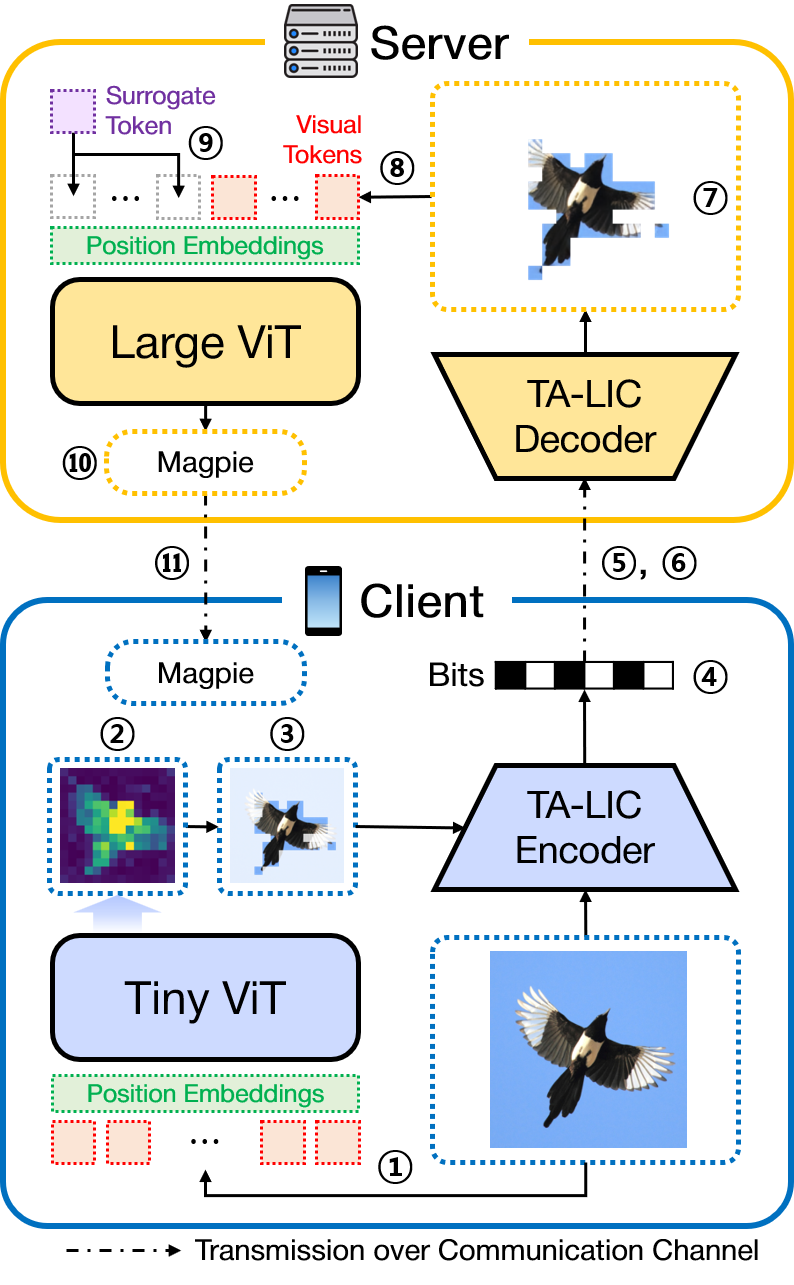}
        \caption{Overview of the proposed token-oriented semantic communication framework based on pretrained ViT models. 
        The circled numbers in the figure follow the same order as the line numbers in Algorithm~\ref{algo:proposed}. 
        }
        \label{fig:overall_framework}
        \vspace{-4mm}
    \end{figure}

    This section presents the client--server collaborative inference procedure of the proposed token-oriented semantic communication framework, deferring its technical components to Section~\ref{sec:main}. 
    We consider a setting in which a resource-constrained client cooperates with a computationally capable server. 
    The client employs a lightweight ViT-based classifier (e.g., DeiT-Tiny~\cite{Touvron2021training}) together with a token-aligned LIC (TA-LIC) encoder, whereas the server employs a large-scale ViT classifier (e.g., DeiT-III-Large~\cite{Touvron2022Deit}) and the corresponding TA-LIC decoder.     

    Our overall framework is illustrated in Fig.~\ref{fig:overall_framework}, where the circled numbers follow the same order as the line numbers in Algorithm~\ref{algo:proposed}. 
    First, the client extracts visual tokens $\mathbf{X}^t$ from the input image $\mathbf{X}$ and computes the token-level task relevance $\bar{\mathbf{r}}$ using the lightweight ViT. 
    Based on these scores, the client selects the task-relevant visual tokens and obtains a token-selection map $\mathbf{s} \in \{0,1\}^{N}$, where $N$ denotes the number of visual tokens. 
    The client then generates the compressed bitstream $\mathbf{b}$, which contains the visual latent representations corresponding to the positions of the selected tokens and the hyperpriors, and transmits $(\mathbf{b}, \mathbf{s})$ to the server. 
    After transmission, the server receives $(\widehat{\mathbf{b}}, \widehat{\mathbf{s}})$ and reconstructs the image $\widehat{\mathbf{X}}$, which is then converted into server-side visual tokens $\widehat{\mathbf{X}}^{t}$.
    From $\widehat{\mathbf{X}}^{t}$, the input sequence $\widetilde{\mathbf{X}}^{t}$ of the server-side ViT is constructed via surrogate token substitution.
    Finally, the server-side predictor produces the classification result $y$ and returns it to the client.
    The overall inference procedure is described in Algorithm~\ref{algo:proposed}. 

    \begin{algorithm}[t]    
    \caption{Inference Procedure of the Proposed Framework}
    \label{algo:proposed}
    \textbf{Input:} Input image $\mathbf{X}$, token-selection threshold $\delta$, client tokenizer $f_{\alpha}: \mathbf{X} \rightarrow \mathbf{X}^{t}$, server tokenizer $f_{\beta}: \widehat{\mathbf{X}} \rightarrow \widehat{\mathbf{X}}^{t}$, client classifier $f_{\theta}: \mathbf{X}^{t} \rightarrow \bar{\mathbf{r}}$, server classifier $f_{\phi}: \widetilde{\mathbf{X}}^{t} \rightarrow y$, TA-LIC encoder $f_{\psi}: (\mathbf{X}, \mathbf{s}) \rightarrow \mathbf{b}$, and TA-LIC decoder $f_{\xi}: (\widehat{\mathbf{b}}, \widehat{\mathbf{s}}) \rightarrow \widehat{\mathbf{X}}$. \\
    \textbf{Output:} Classification result $y$.
    \begin{algorithmic}[1]
        \State $\mathbf{X}^{t} \leftarrow f_{\alpha}(\mathbf{X})$ \Comment{Client-side visual token extraction}
        \State $\bar{\mathbf{r}} \leftarrow f_{\theta}(\mathbf{X}^{t})$ \Comment{Token-level task relevance}
        \State Obtain token-selection map $\mathbf{s} \in \{0,1\}^{N}$ from $\bar{\mathbf{r}}$ using threshold $\delta$
        \State $\mathbf{b} \leftarrow f_{\psi}(\mathbf{X}, \mathbf{s})$ \Comment{Compressed bitstream}
        \State Transmit $(\mathbf{b}, \mathbf{s})$ over the communication channel
        \State Server receives $(\widehat{\mathbf{b}}, \widehat{\mathbf{s}})$
        \State $\widehat{\mathbf{X}} \leftarrow f_{\xi}(\widehat{\mathbf{b}}, \widehat{\mathbf{s}})$ \Comment{Reconstructed image}
        \State $\widehat{\mathbf{X}}^{t} \leftarrow f_{\beta}(\widehat{\mathbf{X}})$ \Comment{Server-side visual token extraction}
        \State Construct $\widetilde{\mathbf{X}}^{t}$ from $\widehat{\mathbf{X}}^{t}$ via surrogate token substitution % short version
        \State $y \leftarrow f_{\phi}(\widetilde{\mathbf{X}}^{t})$ \Comment{Final inference result}
        \State Return $y$ to the client
    \end{algorithmic}
    \end{algorithm}

    We integrate token selection into LIC by retaining only the visual latent representations corresponding to the task-relevant token positions.
    Consequently, the latent payload is reduced according to the number of selected tokens.
    The transmitted bitstream also includes the hyperpriors and the selection map $\mathbf{s}$, but the bits allocated to them are sufficiently small compared to those used for the visual latent representations. 
    As a result, the bitstream size decreases approximately linearly with the number of selected tokens, as analyzed in Section~\ref{sec:main-lic}.

    \section{Main Technical Components}\label{sec:main}

    In this section, we present three technical components of the proposed framework: (1) token-aligned learned image compression, (2) layer-selective attention rollout, and (3) learnable surrogate token substitution.     
    The client first estimates the token-level task relevance of visual tokens via layer-selective attention rollout on a lightweight ViT model. 
    It then performs token-wise on/off selection based on the estimated relevance, and the LIC model generates a bitstream containing the latent representations at the selected token positions and the hyperpriors. 
    At the server, the reconstructed patches at unselected positions are unreliable because the corresponding latents are not transmitted.
    To prevent these unreliable patches from degrading downstream inference, the server replaces the tokens at those positions with a learnable surrogate token.

    \subsection{Token-Aligned Learned Image Compression}\label{sec:main-lic}

    Given the token-level task relevance $\bar{\mathbf{r}}$ estimated from the attention weights of the client-side ViT (Section~\ref{sec:main-attention}), the client selects task-relevant tokens to reduce communication overhead.
    A straightforward approach is to transmit the selected image patches directly in the pixel domain~\cite{Im2024Attention} or after importance-aware scalar quantization~\cite{Park2025Vision}.
    However, both approaches encode each patch independently and thus cannot exploit the statistical redundancy that learned image compression captures, remaining far less rate-efficient than LIC models, as demonstrated in Section~\ref{sec:experimental}.
    This motivates integrating token selection into the compression process itself.
    
\begin{figure}[t] 
        \centering
        \includegraphics[width=0.43\textwidth]{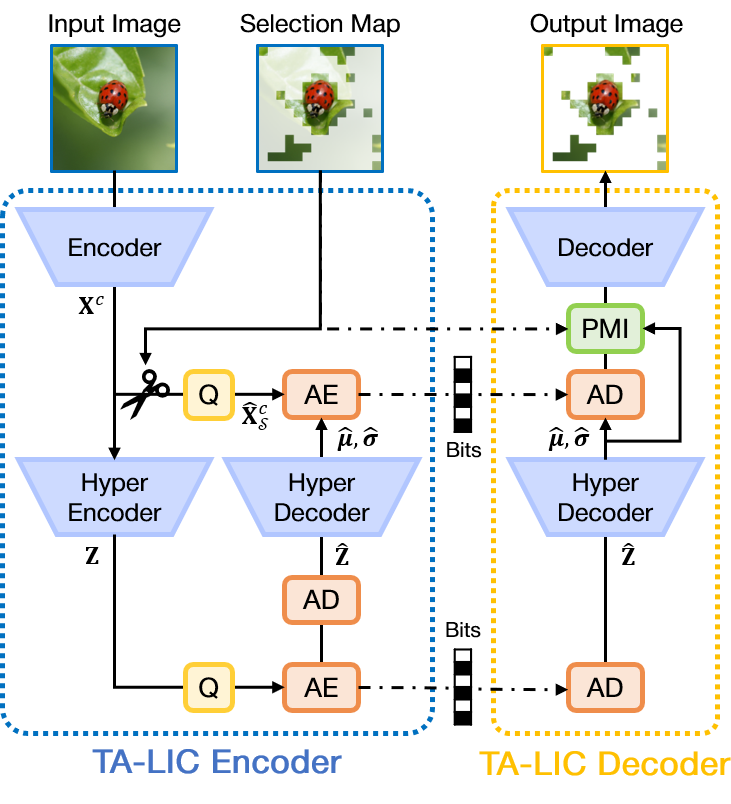}
        \caption{Proposed token-aligned learned image compression (TA-LIC) framework. \emph{Q}, \emph{AE}, \emph{AD}, and \emph{PMI} refer to the quantizer, arithmetic encoder, arithmetic decoder, and predicted-mean imputation block, respectively. At the TA-LIC encoder, only the latent representations at task-relevant token positions are retained and compressed into a bitstream. At the TA-LIC decoder, missing latent positions are imputed with hyperprior-predicted means before image reconstruction by the decoder. Although the reconstructed image retains the original input resolution, only the patches at the positions of the transmitted latents are forwarded to the server-side ViT; the output image is therefore shown with the remaining patches omitted.} 
        \label{fig:lic_framework}
        \vspace{-4mm}
    \end{figure}

    We propose \emph{token-aligned LIC} (TA-LIC), which integrates attention-aware token selection with an LIC model by operating in the latent domain, as shown in Fig.~\ref{fig:lic_framework}.
    The key enabler is a structural alignment between the visual tokens of ViT and the latent representations of LIC: the encoder maps an input image $\mathbf{X}$ to a latent representation $\mathbf{X}^c$, which is quantized to $\widehat{\mathbf{X}}^c \in \mathbb{Z}^{N \times D_c}$ with spatial resolution $\left(\frac{H}{P}, \frac{W}{P}\right)$, yielding $N = \frac{HW}{P^2}$ spatial positions identical to the number of visual tokens.
    Note that both architectures conventionally reduce the input resolution by the same factor $P = 16$: ViT through a patch size of $(16, 16)$~\cite{Dosovitskiy2021image,Touvron2021training,Touvron2022Deit}, and LIC through four stride-2 convolutional stages~\cite{Balle2018Variational,Minnen2018Joint,He2022Elic}.
    This one-to-one correspondence between the $i$th visual token $\mathbf{x}_i^t$ and the $i$th latent vector $\widehat{\mathbf{x}}_i^c$, denoting the $i$th rows of $\mathbf{X}^t$ and $\widehat{\mathbf{X}}^c$, respectively, allows the token-selection map $\mathbf{s} \in \{0,1\}^{N}$ to directly index the latent representations, controlling which latent vectors are included in the transmitted bitstream without modifying the LIC architecture.

    Given the task relevance $\bar{\mathbf{r}}\in \mathbb{R}^{N}$, we construct the selection map $\mathbf{s}$ using the attention-sum threshold selection of~\cite{Im2024Attention}, where $s_i = 1$ indicates that the $i$th token is selected.
    Tokens are selected in descending order of relevance until the cumulative sum of relevance first exceeds a preset threshold $\delta$, so that the number of selected tokens varies across images according to semantic content, such as object size and inference difficulty.
    We denote the resulting sets of selected and unselected spatial indices as $\mathcal{S} = \{i \mid s_i = 1\}$ and $\mathcal{U} = \{i \mid s_i = 0\}$, respectively.
    
    \subsubsection{Selective Latent Transmission}
    
    To selectively include only task-relevant latent representations in the transmitted bitstream, we exploit the structure of the hyperprior-based entropy model~\cite{Balle2018Variational}.
    Specifically, the distribution of the $d$th element of the $i$th latent $\widehat{x}^c_{i,d}$ is fully parameterized by the predicted mean $\mu_{i,d}$ and scale $\sigma_{i,d}$, obtained from the hyper decoder $h_s$ using the quantized hyperprior $\widehat{\mathbf{Z}}$~\cite{Minnen2018Joint}.
    Since the latent elements are modeled as conditionally independent given $\widehat{\mathbf{Z}}$, each element can be entropy-coded and decoded using only its own predicted parameters $(\mu_{i,d}, \sigma_{i,d})$, without access to any other latent element.
    Consequently, the TA-LIC encoder guides the arithmetic encoder to encode only the latent vectors at positions $i\in\mathcal S$, bypassing entropy coding for positions $i\in\mathcal{U}$ entirely.
    
    The transmitted bitstream comprises the tuple $\left(\widehat{\mathbf{X}}^c_{\mathcal S},\, \widehat{\mathbf{Z}},\, \mathbf{s}\right)$, where $\widehat{\mathbf{X}}^c_{\mathcal S} \in \mathbb{Z}^{|\mathcal S| \times D_c}$ stacks the retained latent vectors $\widehat{\mathbf{x}}^c_i$, $i \in \mathcal{S}$, in ascending order of spatial index.
    This selective encoding reduces the communication cost; the resulting rate reduction is quantified in the theoretical analysis below.
    
    \subsubsection{Theoretical Analysis of Communication Efficiency} \label{sec:main-lic-analysis}
    
    To analyze the communication efficiency, we establish a theoretical bound demonstrating that an approximately linear reduction in bit rate is achievable.
    The analysis assumes an ideal entropy coder in which the code length of each symbol approaches its self-information.
    
    The number of bits required to encode the retained latents is given by 
    \begin{equation} \label{eq:bitrate-of-selected}
        B_{\widehat{\mathbf{X}}^c_{\mathcal S}}
        = \left\lceil - \sum_{i \in \mathcal{S}} \sum_{d}
          \log_2 p\!\left(\widehat{x}^c_{i,d} \mid \widehat{\mathbf{Z}}\right)
          \right\rceil.
    \end{equation}
    Analogously, the number of bits required to encode the complete latent matrix $\widehat{\mathbf{X}}^c$, i.e., over both selected and unselected positions, is
    \begin{equation} \label{eq:bitrate-of-full}
        B_{\widehat{\mathbf{X}}^c}
        = \left\lceil - \sum_{i=1}^N \sum_{d}
          \log_2 p\!\left(\widehat{x}^c_{i,d} \mid \widehat{\mathbf{Z}}\right)
          \right\rceil.
    \end{equation}
    
    Transmitting the complete latent representation together with the side information would then require
    \begin{equation} \label{eq:total-rate}
        B_\mathrm{tot} = B_{\widehat{\mathbf{X}}^c} + B_{\widehat{\mathbf{Z}}},
    \end{equation}
    where $B_{\widehat{\mathbf{Z}}}$ denotes the number of bits allocated to the quantized hyperprior.
    As the proposed framework transmits only the retained latents, however, the actual number of transmitted bits is
    \begin{equation} \label{eq:transmitted-rate}
        B = B_{\widehat{\mathbf{X}}^c_{\mathcal S}} + B_{\widehat{\mathbf{Z}}} + B_{\mathbf{s}},
    \end{equation}
    where $B_{\mathbf{s}}$ denotes the number of bits allocated to the binary selection map.
    
    Consequently, the ratio of the actual to the total bits admits the upper bound
    \begin{equation} \label{eq:reduced-ratio}
        \frac{B}{B_\mathrm{tot}}
        \le \eta + \frac{\rho}{\rho + (1-\rho)k},
        \quad \text{where } k = \frac{\bar{H}_{\mathcal{U}}}{\bar{H}_{\mathcal{S}}},
    \end{equation}
    where $\rho = \frac{|\mathcal{S}|}{|\mathcal{S}|+|\mathcal{U}|}$ denotes the token selection ratio, $\eta = \frac{B_{\widehat{\mathbf{Z}}} + B_{\mathbf{s}}}{B_\mathrm{tot}}$ denotes the side-information overhead ratio including the hyperprior and the selection map, and $\bar{H}_{\mathcal{S}}$ and $\bar{H}_{\mathcal{U}}$ denote the average self-information per latent over the selected and unselected positions, respectively.
    
    In the regime $k \approx 1$, where the selected and unselected regions exhibit comparable information density, the bound simplifies to
    \begin{equation} \label{eq:approx-bit}
        \frac{B}{B_\mathrm{tot}} \lesssim \eta + \rho,
    \end{equation}
    showing that the proposed selective transmission yields a rate reduction approximately linear in the selection ratio $\rho$, offset only by the side-information overhead $\eta$.
    The bound further characterizes the nonlinear regimes: the reduction is sublinear when the unselected region is information-sparse ($k < 1$), and superlinear when it is information-dense relative to the selected region ($k > 1$).

    \subsubsection{Predicted-Mean Imputation}

    Before latent decoding, the TA-LIC decoder imputes the unselected latent positions with their predicted means, derived from the received hyperprior $\widehat{\mathbf{Z}}$ following the mean-scale hyperprior architecture~\cite{Minnen2018Joint}.
    From a statistical perspective, the predicted mean is the conditional expectation of the local latent feature given the context captured by the hyperprior, and is thus a principled substitute for the missing latent.
    This choice matters because the decoder uses convolutional layers with overlapping receptive fields: the reconstruction of each selected patch is influenced by the latents at neighboring positions, so the imputed values directly affect the quality of the neighboring selected patches.
    Substituting the unselected latents with their conditional expectations preserves the local statistical consistency of the latent map, improving downstream classification accuracy over zero padding, as shown in Section~\ref{sec:experimental-lic}.
    
    Furthermore, since $\widehat{\boldsymbol{\mu}} \in \mathbb{R}^{N \times D_c}$ is already produced by the hyper decoder for entropy decoding, the imputation introduces negligible additional computational overhead.
    The $i$th row of the imputed latent representation $\widetilde{\mathbf{X}}^c$ at the TA-LIC decoder is
    \begin{equation} \label{eq:mean-imputation}
    \widetilde{\mathbf{x}}^c_i = s_i \, \widehat{\mathbf{x}}^c_i + (1 - s_i) \, \widehat{\boldsymbol{\mu}}_i, \quad i = 1, \dots, N,
    \end{equation}
    where $\widehat{\boldsymbol{\mu}}_i$ denotes the $i$th row of $\widehat{\boldsymbol{\mu}}$. 
    Note that predicted-mean imputation is introduced solely to improve the reconstruction of the selected image patches: the reconstructed patches at unselected positions are not forwarded to the server-side predictor, as described in Section~\ref{sec:main-surrogate}.

    \subsection{Layer-Selective Attention Rollout} \label{sec:main-attention}

    In transformer architectures, token interactions are primarily mediated by multi-head self-attention, whose attention weights capture global dependencies among tokens.
    For image classification with ViTs, the final prediction is produced from the class token, so the attention from the class token to the visual tokens naturally indicates how much each visual token contributes to the classification task. 

    \begin{figure}[t] 
        \centering
        \includegraphics[width=0.48\textwidth]{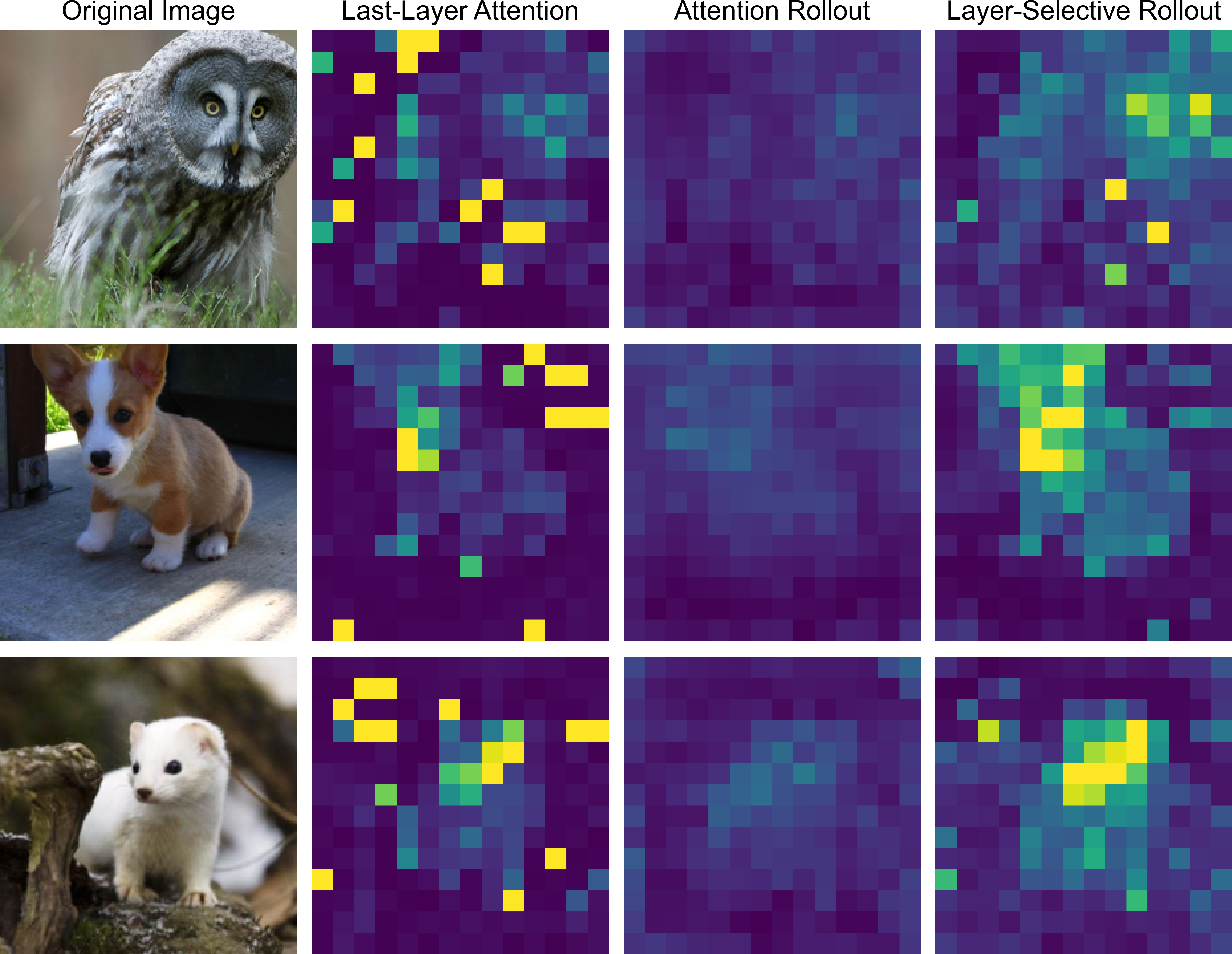}
        \caption{Visualization of token-level task relevance from DeiT-Tiny~\cite{Touvron2021training}. The first column shows example ImageNet images labeled as `great gray owl,' `Pembroke,' and `polecat,' respectively. The remaining columns, from left to right, show task-relevance maps obtained using last-layer attention, attention rollout, and layer-selective attention rollout from the 7th to the 12th layers, respectively.}
        \label{fig:task_relevance}
        \vspace{-4mm}
    \end{figure}
    
    Along this line, last-layer class-token attention has been widely used for task relevance estimation owing to its simplicity~\cite{Liu2023efficient,Liu2024adaptive,Im2024Attention,Park2025Vision}.
    However, as shown in Fig.~\ref{fig:task_relevance}, last-layer attention often assigns high relevance to non-object regions, overestimating the task relevance of background patches.
    This limitation may arise because visual tokens are progressively contextualized across transformer layers~\cite{Abnar2020quantifying,Chefer2021transformer,Chefer2021Generic}: by the final layer, each token aggregates information from other tokens and no longer represents only its own patch, so the class-token attention at a single layer does not faithfully reflect which patches originally contributed to the prediction.

    Attention rollout was originally proposed to trace attention propagation among language tokens in natural language processing~\cite{Abnar2020quantifying}.
    To account for the progressive contextualization described above, it multiplies the attention matrices across all layers, thereby explicitly tracking the information flow from the input to the final representation.
    As shown in Fig.~\ref{fig:task_relevance}, the resulting map is largely concentrated on the main object.
    However, when the selected patches are transmitted for server-side inference, attention rollout yields lower classification accuracy than last-layer attention unless only a small fraction of patches is retained~\cite{Im2024Attention}, indicating that accurate object localization does not necessarily translate into accurate task-relevance estimation.

    \begin{figure}[t]
        \centering
        \includegraphics[width=0.48\textwidth]{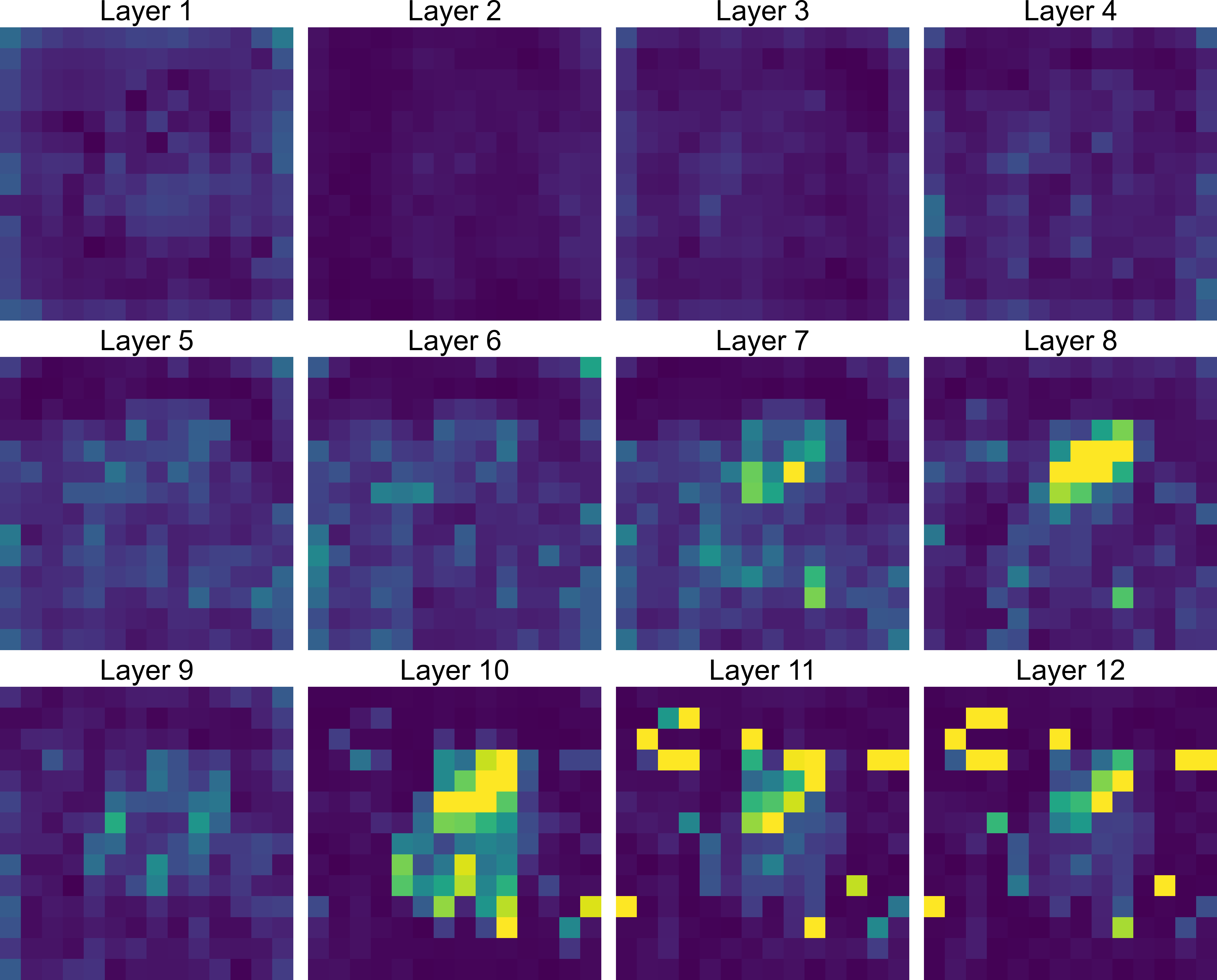}
        \caption{Visualization of class-token attention weights across all layers of DeiT-Tiny. The original image is `polecat,' same as Fig.~\ref{fig:task_relevance}. Note that the visualized attentions are averaged along the head dimension.}
        \label{fig:vis-attention}
        \vspace{-4mm}
    \end{figure}

    To understand why both methods fall short, we visualize the class-token attention weights across all layers of DeiT-Tiny~\cite{Touvron2021training} in Fig.~\ref{fig:vis-attention}.
    The attention patterns are noisy and weakly interpretable in the early layers, concentrate on the main object in the middle layers, and drift toward subsets of the background in the final layers.
    This layer-dependent behavior explains the preceding observations: last-layer attention inherits the background drift of the final layers, whereas attention rollout accumulates the noise of the early layers.
    These observations motivate aggregating attention over a selected range of layers that excludes the noisy early layers, rather than relying on a single layer or on all layers.

    Therefore, we propose \emph{layer-selective attention rollout}, which aggregates attention weights over a selected range of ViT layers.
    The layer-selective attention rollout from the $L_s$th layer to the $L_e$th layer, where $L_s \leq L_e$, is computed as
    \begin{equation}
        \widetilde{\mathbf{A}}_{l} = \mathbb{E}_h \left[ \mathbf{A}_{l,h} \right] + \mathbf{I}_{N+1}, \quad
        \mathbf{R} = \widetilde{\mathbf{A}}_{L_e} \widetilde{\mathbf{A}}_{L_e-1} \cdots \widetilde{\mathbf{A}}_{L_s},
    \end{equation}
    and the token-level relevance $\bar{\mathbf{r}}$ is then obtained via~\eqref{eq:relevance-normalization}.
    For DeiT-Tiny with $L=12$ layers, we set $(L_s, L_e) = (7, 12)$, determined by the layer-range study in Section~\ref{sec:experimental-attention}.
    The last layer is included because it directly shapes the class token consumed by the classification head~\cite{Dosovitskiy2021image}, while the noisy early layers are excluded.
    Examples of the resulting relevance maps are shown in Fig.~\ref{fig:task_relevance}, assigning relevance to the main object together with parts of its surrounding regions.

    As shown in Section~\ref{sec:experimental-attention}, layer-selective attention rollout estimates task relevance more accurately than last-layer attention and attention rollout, and matches or exceeds gradient-weighted attention~\cite{Chefer2021Generic} across our main operating region.
    Achieving this accuracy without gradients is what matters in practice: gradient-weighted attention requires a backward pass through the client-side ViT at every inference, and the additional computation and peak-memory footprint for storing intermediate activations are costly on resource-constrained clients.
    Layer-selective attention rollout attains this accuracy, exceeding gradient-weighted attention on our main client configuration, from a single forward pass, making precise task-relevance estimation practical on such clients.

    \subsection{Surrogate Token Substitution}\label{sec:main-surrogate}

    Through the TA-LIC decoder, the server obtains a reconstructed image $\widehat{\mathbf{X}} \in \mathbb{R}^{H \times W \times C}$ that retains the original spatial resolution, in which the patches at unselected positions are synthesized from the hyperprior-predicted means and thus reflect only a coarse statistical estimate of the original content at those positions.
    To prevent these patches from serving as misleading visual evidence for the server-side predictor, we introduce a single learnable token, termed the \emph{surrogate token}, that replaces the tokens at unselected positions and is optimized while the pretrained backbone is kept frozen.
    The modified input token sequence $\widetilde{\mathbf{X}}^t \in \mathbb{R}^{(N+1) \times D}$ is constructed by prepending the class token $\mathbf{x}_\mathrm{cls}$ and assigning the surrogate token $\mathbf{x}_\mathrm{sur} \in \mathbb{R}^{1 \times D}$ to the unselected positions:
    \begin{align} \label{eq:surrogate-substitution}
        \widetilde{\mathbf{x}}^t_1 &= \mathbf{x}_\mathrm{cls}, \\
        \widetilde{\mathbf{x}}^t_{i+1} &= s_i \, \widehat{\mathbf{x}}^t_i + (1 - s_i) \, \mathbf{x}_\mathrm{sur} + \mathbf{e}_i, \quad i = 1, \dots, N,
    \end{align}
    where $\widetilde{\mathbf{x}}^t_i$ and $\mathbf{e}_i$ denote the $i$th row of the resulting token sequence $\widetilde{\mathbf{X}}^t$ and the position embedding $\mathbf{E}$ in~\eqref{eq:vit-input_sequence}, respectively.
    We optimize $\mathbf{x}_\mathrm{sur}$ for the downstream classification objective with the backbone frozen, so that the surrogate token learns to indicate the absence of information rather than supply misleading visual evidence.

    Without any adaptation, the server can perform inference directly on the reconstructed image by replacing $\mathbf{X}^t$ in~\eqref{eq:vit-input_sequence} with $\widehat{\mathbf{X}}^t$ (\emph{direct-reconstruction inference}), or exploit the variable-length input property of ViTs to use only the reconstructed patches at the selected positions (\emph{selected-patch inference}).
    While some accuracy reduction is unavoidable when inferring from partial visual information, both adaptation-free baselines leave the frozen predictor unadapted to the incomplete input and thus miss the accuracy that adaptation could recover.
    At the opposite extreme, retraining the server-side ViT on incomplete images recovers this margin but is computationally expensive~\cite{Hu2021Lora,Jia2022Visual,Yu2024Visual}, and updating the backbone weights can degrade performance on complete or near-complete inputs, where the pretrained DeiT-III-Large~\cite{Touvron2022Deit} already attains its peak accuracy of \SI{86.81}{\%} on ImageNet.
    
    Surrogate token substitution occupies the middle ground between these extremes: the cost of adaptation is a single learnable vector of $D$ parameters ($D = 1024$ for DeiT-III-Large), and since the backbone remains intact, the predictor's behavior on complete inputs is identical to that of the original pretrained model.
    Section~\ref{sec:experimental-surrogate} demonstrates that it maintains downstream accuracy more effectively than both adaptation-free baselines.
    
    Surrogate token substitution further improves the robustness of server-side inference over unreliable channels.
    When the received bitstream differs from the transmitted one because of uncorrected channel impairments, a subset of the transmitted LIC latents may be erased.
    In this case, the server can replace the tokens at the affected positions with the surrogate token, reducing the impact of erasures on classification accuracy; this robustness gain is evaluated over a packet-erasure channel in Section~\ref{sec:experimental-erasure}.

    \section{Experimental Results}\label{sec:experimental}

    We evaluate the proposed token communication framework on ImageNet classification.
    After describing the experimental settings (Section~\ref{sec:experimental-settings}), we first present the main results: the rate--accuracy trade-off against prior token-oriented semantic communication baselines, hand-crafted codecs, and LIC frameworks (Section~\ref{sec:experimental-main}), the robustness of surrogate token substitution under a packet-erasure channel (Section~\ref{sec:experimental-erasure}), and the additional gain of entropy-aware image transmission (Section~\ref{sec:experimental-entropy}).
    We then conduct ablation studies that isolate the contribution of each technical component: token-aligned LIC (Section~\ref{sec:experimental-lic}), layer-selective attention rollout (Section~\ref{sec:experimental-attention}), and surrogate token substitution (Section~\ref{sec:experimental-surrogate}).

    \subsection{Experimental Settings}\label{sec:experimental-settings}

    \begin{table} [t]
    \renewcommand{\arraystretch}{1.2}
        \centering
        \caption{Model Complexity and Classification Accuracy on ImageNet-1k~\cite{Deng2009Imagenet,Balle2018Variational,Touvron2021training,Touvron2022Deit}}
        \begin{tabular}[width=0.2\textwidth]{|c|c|c|c|c|}
        \hline
            Model & Parameters  & Memory & 
            MACs & Classification\\ & (million) & (MB) & (G) & Accuracy (\%)\\\hline \hline
            DeiT-Tiny& 5.7 & 22.9 & 1.26 & 72.2\\
            DeiT-Small& 22.1 & 88.2 & 4.61 & 79.8\\
            DeiT-Base& 86.6 & 346.3 & 17.58 & 81.8\\
            DeiT-III-Large& 304.4 & 1217.5 & 61.6 & 86.8\\ \hline
            TA-LIC encoder& 5.5 & 22.1 & 3.21 & - \\
            TA-LIC decoder& 4.5 & 18.0 & 2.76 & - \\ \hline
        \end{tabular}
        \label{tab:model}
    \end{table}

    In the proposed system, we assume that the client deploys DeiT-Tiny~\cite{Touvron2021training}, while the server employs DeiT-III-Large~\cite{Touvron2022Deit}, reflecting the disparity in memory and computational capacity between the client and the server.
    As shown in Table~\ref{tab:model}, the accuracy gap between the two models is substantial: DeiT-III-Large attains \SI{86.8}{\%}, whereas DeiT-Tiny reaches only \SI{72.2}{\%}.
    However, this accuracy improvement comes with substantially higher resource requirements, including approximately $50\times$ larger memory consumption and computational complexity than those of DeiT-Tiny.
    In contrast, the TA-LIC encoder based on the hyperprior architecture~\cite{Balle2018Variational} requires only \SI{22.1}{MB} of memory, making it suitable for resource-constrained clients.
    Overall, the client-side models, consisting of DeiT-Tiny and the TA-LIC encoder, require only \SI{45}{MB} of memory and \SI{4.47}{GMACs}, whereas the server-side models, consisting of DeiT-III-Large and the TA-LIC decoder, require more than \SI{1.2}{GB} of memory and approximately \SI{64.36}{GMACs}.

    We evaluate the framework on ImageNet~\cite{Deng2009Imagenet} for image classification.
    Each image is center-cropped and resized to a resolution of $(224, 224)$ pixels.
    We adopt publicly available pretrained ViT parameters, where DeiT-III-Large is additionally pretrained on ImageNet-21k~\cite{Touvron2022Deit}.
    The LIC model and the surrogate token are trained separately, with the settings described below.

    \begin{itemize}
        \item \emph{LIC training}:
        We train a hyperprior model~\cite{Balle2018Variational} with a mean hyperprior~\cite{Minnen2018Joint} for image reconstruction on ImageNet, setting the channel dimensions of the latent representation and the hyperprior to $192$ and $128$, respectively.
        A separate LIC model is trained for each considered value of $\lambda$, while the remaining optimization settings follow~\cite{Balle2018Variational,Minnen2018Joint}.

        \item \emph{Surrogate token training}:
        Following Section~\ref{sec:main-surrogate}, we optimize a single surrogate token $\mathbf{x}_{\mathrm{sur}} \in \mathbb{R}^{1 \times D}$ for the classification objective while keeping the pretrained backbones frozen.
        Surrogate token training emulates the proposed pipeline: the client selects task-relevant tokens via layer-selective attention rollout under threshold $\delta_{\mathrm{sur}}$ and transmits a bitstream compressed by an LIC model pretrained with $\lambda_{\mathrm{sur}}$, after which the server replaces the unselected positions with the surrogate token before inference.
        The token is trained for $20$ epochs, with the remaining settings following~\cite{He2022masked}.
    \end{itemize} 

    The selection map $\mathbf{s} \in \{0, 1\}^{N}$ accompanies the encoded image data $(\widehat{\mathbf{X}}^c_{\mathcal{S}}, \widehat{\mathbf{Z}})$ so that the server can both reconstruct the image and identify the positions to be replaced by the surrogate token.
    At our resolution, the selection map occupies only $N = 196$ bits per image, corresponding to less than $0.004$ bits per pixel (\SI{}{bpp}), where the bit rate is computed as the average of $\frac{\text{sent bits}}{\text{number of pixels}}$ over the evaluation set.
    We therefore omit the cost of $\mathbf{s}$ when computing the overall bit rate.

    \subsection{Rate--Accuracy Trade-Off}\label{sec:experimental-main}

\begin{figure}[t]
        \centering
        \subfloat[Token-oriented semantic communication frameworks]{%
            \includegraphics[width=0.4\textwidth]{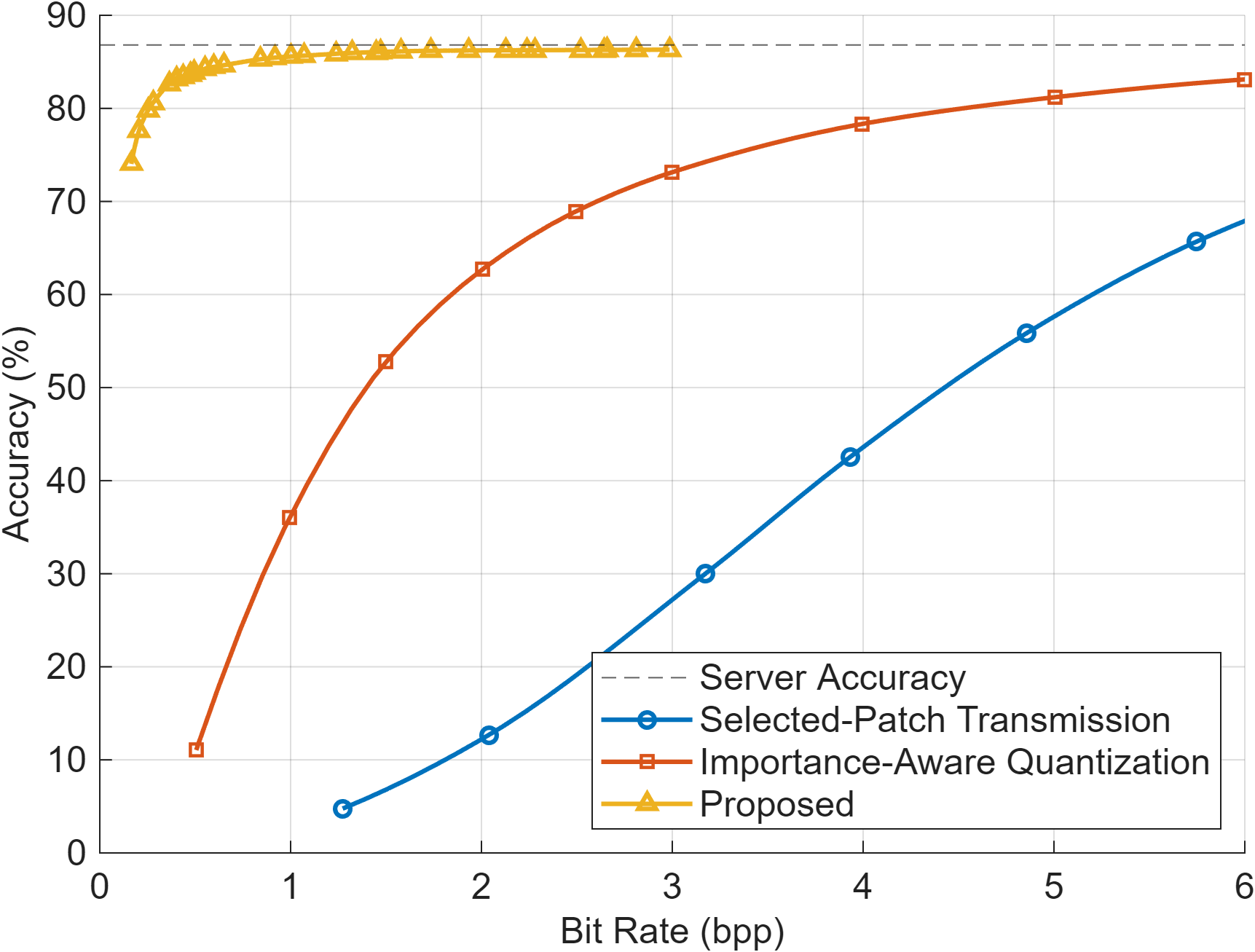}%
            \label{fig:exp-main-token_oriented_frameworks}}
        \vfil
        \subfloat[Hand-crafted codecs]{%
            \includegraphics[width=0.4\textwidth]{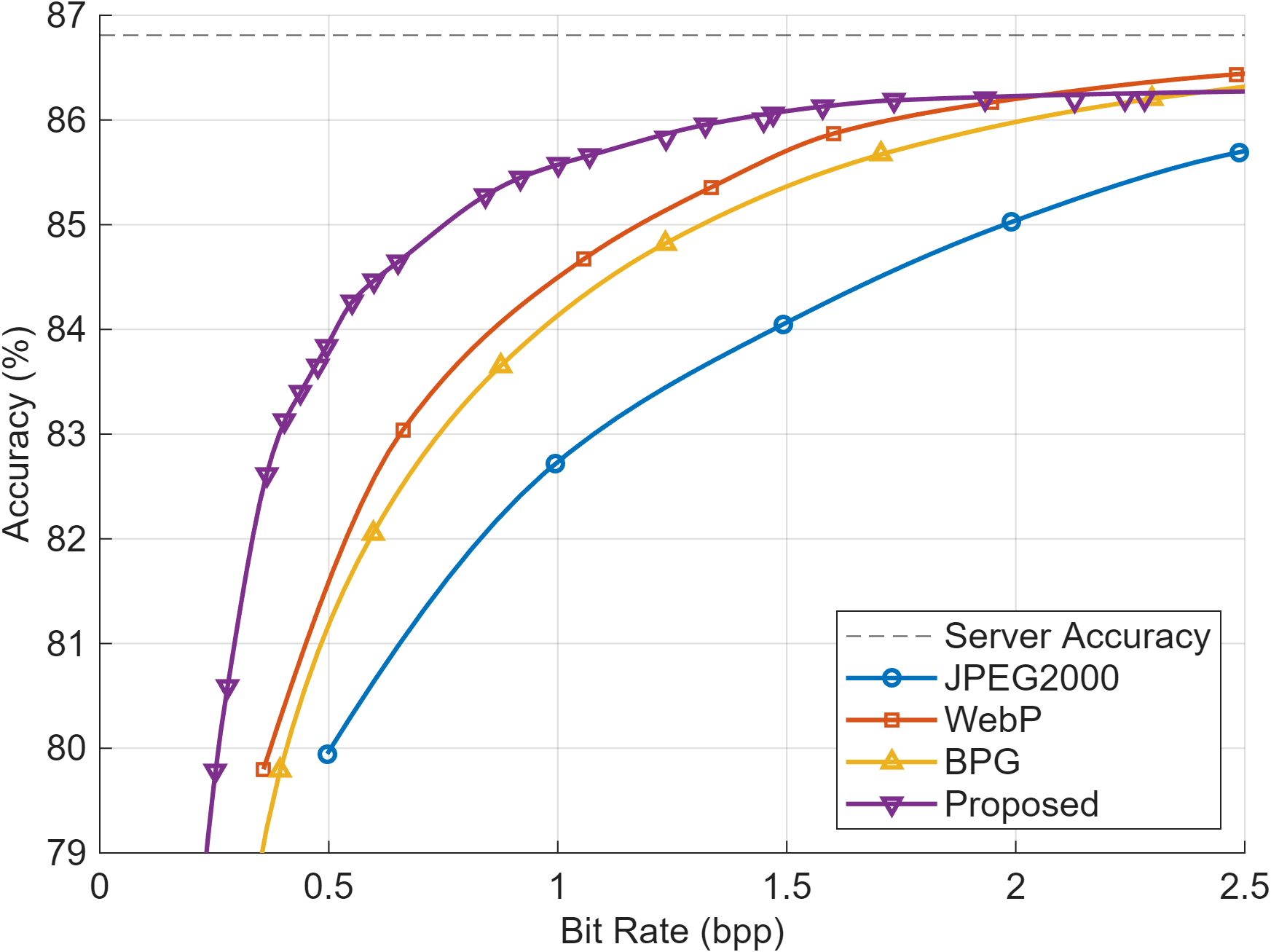}%
            \label{fig:exp-main-hand_codecs}}
        \vfil
        \subfloat[LIC frameworks]{%
            \includegraphics[width=0.4\textwidth]{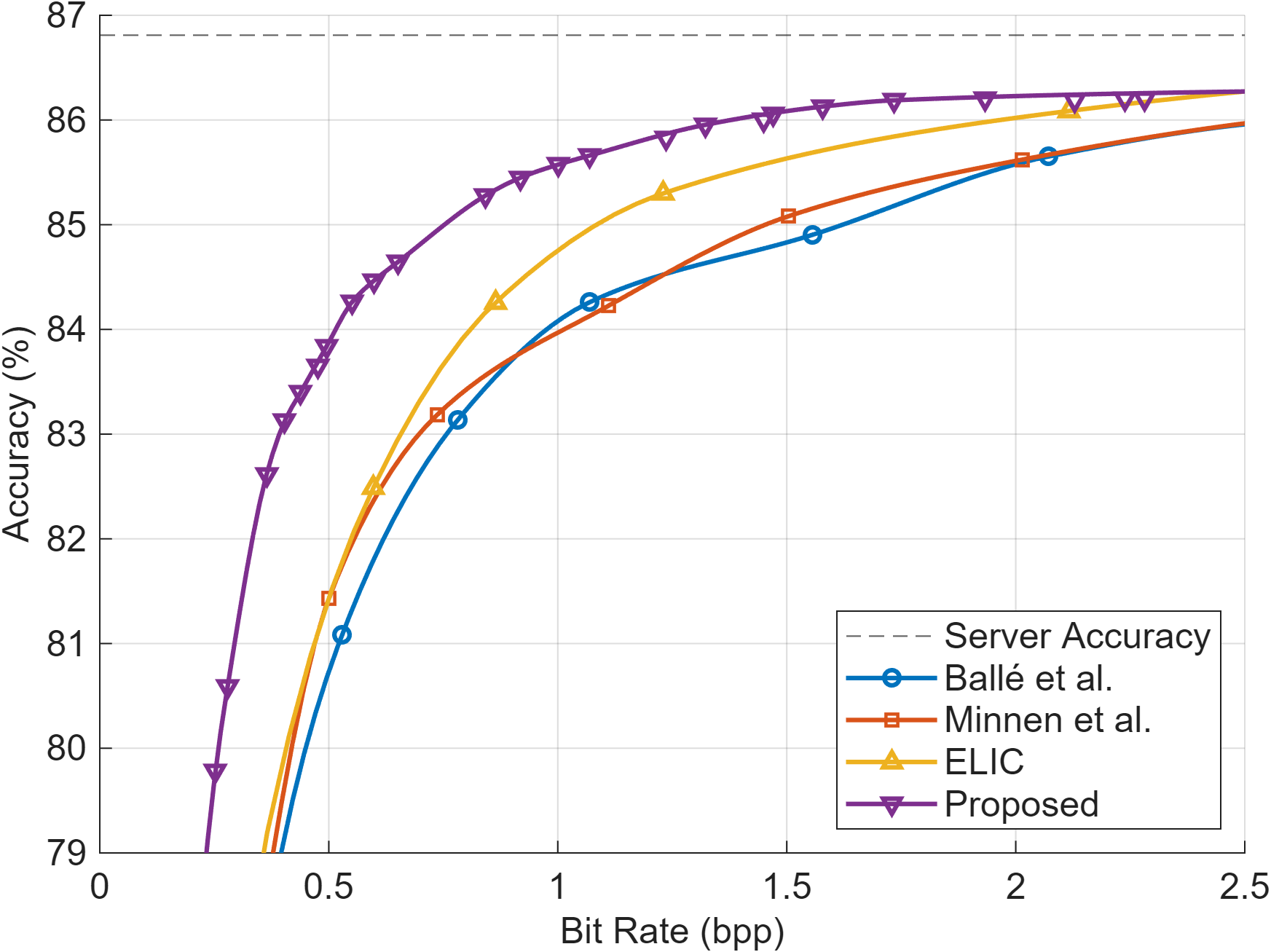}%
            \label{fig:exp-main-LIC}}
        \caption{Trade-off between bit rate and classification accuracy of the proposed framework against (a) prior token-oriented semantic communication frameworks~\cite{Im2024Attention,Park2025Vision}, (b) hand-crafted codecs, and (c) LIC frameworks~\cite{Balle2018Variational,Minnen2018Joint,He2022Elic}. \emph{Server accuracy} corresponds to the original downstream performance of DeiT-III-Large, i.e., \SI{86.81}{\%}, attainable under lossless transmission at \SI{24}{bpp}. \emph{Proposed} denotes the achievable curve formed by connecting the Pareto-optimal operating points of the proposed framework.}
        \label{fig:exp-main}
        \vspace{-4mm}
    \end{figure}

    In Fig.~\ref{fig:exp-main}, we evaluate the trade-off between bit rate and classification accuracy of the proposed framework against three baseline families: prior token-oriented semantic communication frameworks, hand-crafted codecs, and LIC models.
    For a fair comparison, all baselines decode the transmitted representation into an image that is classified by the same pretrained server model, DeiT-III-Large, fixing the upper-bound accuracy under lossless transmission to \SI{86.81}{\%}; the rate--accuracy curves thus differ only in how efficiently each method compresses and transmits the image.
    End-to-end frameworks, in contrast, jointly train a task-specific classifier with the transceiver, yielding a different accuracy ceiling and no standalone image, and thus are not directly comparable under this controlled protocol.
    
    The rate--accuracy trade-off of the proposed framework is characterized in the token-aligned LIC module by varying the token-selection threshold $\delta$ and switching among LIC models trained with different values of $\lambda$, where $\delta$ controls the number of selected tokens and $\lambda$ determines the operating point of the LIC model.
    We sweep $\delta$ from $0.5$ to $0.98$ and $\lambda$ from $0.0075$ to $2$, where larger values of either generally yield higher bit rates and improved downstream performance.
    The remaining components are fixed during inference: $(L_s, L_e) = (7, 12)$ for layer-selective attention rollout, and a surrogate token trained with $(\delta_{\mathrm{sur}}, \lambda_{\mathrm{sur}}) = (0.35, 0.03)$.
    The effects of $(L_s, L_e)$ and $(\delta_{\mathrm{sur}}, \lambda_{\mathrm{sur}})$ are further discussed in Section~\ref{sec:experimental-attention} and Section~\ref{sec:experimental-surrogate}, respectively.

    As shown in Fig.~\ref{fig:exp-main}\subref{fig:exp-main-token_oriented_frameworks}, the proposed framework achieves higher downstream accuracy at comparable bit rates to the token-oriented baselines, namely selected-patch transmission~\cite{Im2024Attention} and importance-aware quantization~\cite{Park2025Vision}.
    Since these baselines also exploit task relevance but transmit in the pixel domain, this gain comes mainly from the rate efficiency of token-aligned LIC.
    Against hand-crafted codecs (JPEG2000, WebP, and BPG), Fig.~\ref{fig:exp-main}\subref{fig:exp-main-hand_codecs} shows a favorable trade-off below \SI{2}{bpp}: the proposed framework attains \SI{85.83}{\%} accuracy at \SI{1.24}{bpp}, only $0.98$ percentage points below the server accuracy, whereas WebP and BPG require about \SI{1.57}{bpp} and \SI{1.82}{bpp} to match it.
    A comparable advantage holds against task-agnostic LIC frameworks~\cite{Balle2018Variational,Minnen2018Joint,He2022Elic}, where Fig.~\ref{fig:exp-main}\subref{fig:exp-main-LIC} shows higher accuracy at comparable bit rates below \SI{2.5}{bpp}.
    Although the proposed selective transmission is not directly compatible with autoregressive context models such as ELIC~\cite{He2022Elic}, its implementation with the simpler mean-scale hyperprior still achieves a more favorable rate--accuracy trade-off than ELIC. 
    This result highlights the benefit of task-relevance-guided latent selection over improving reconstruction-oriented entropy modeling alone.
    Overall, the gains against hand-crafted codecs and prior LIC frameworks confirm the benefit of incorporating task relevance, beyond optimizing reconstruction fidelity alone.

    \subsection{Classification Accuracy over Erasure Channel}\label{sec:experimental-erasure}

    To evaluate the robustness of surrogate token substitution over unreliable channels, we measure the classification accuracy of the proposed framework on a packet-erasure channel.
    Because the LIC model uses arithmetic coding to compress both latent representations and hyperpriors, a single erasure may disrupt the synchronization of the entropy model between the encoder and decoder; we therefore packetize them and independently encode each packet to localize the impact of erasures.

    The hyperpriors, which provide the side information for latent decoding, are placed in a single packet assumed to be received without error.
    The latents are then greedily packed into the remaining packets in descending order of their per-latent self-information $H_{\widehat{\mathbf{x}}_i^c} = - \sum_{d} \log_2 p(\widehat{x}^c_{i, d} \mid \widehat{\mathbf{Z}})$.
    Each packet, constrained to a maximum transmission unit of \SI{1460}{bytes}, is then independently arithmetic-coded and transmitted over the erasure channel with erasure probability $p_e$.

    \begin{figure}[t] 
        \centering
        \includegraphics[width=0.4\textwidth]{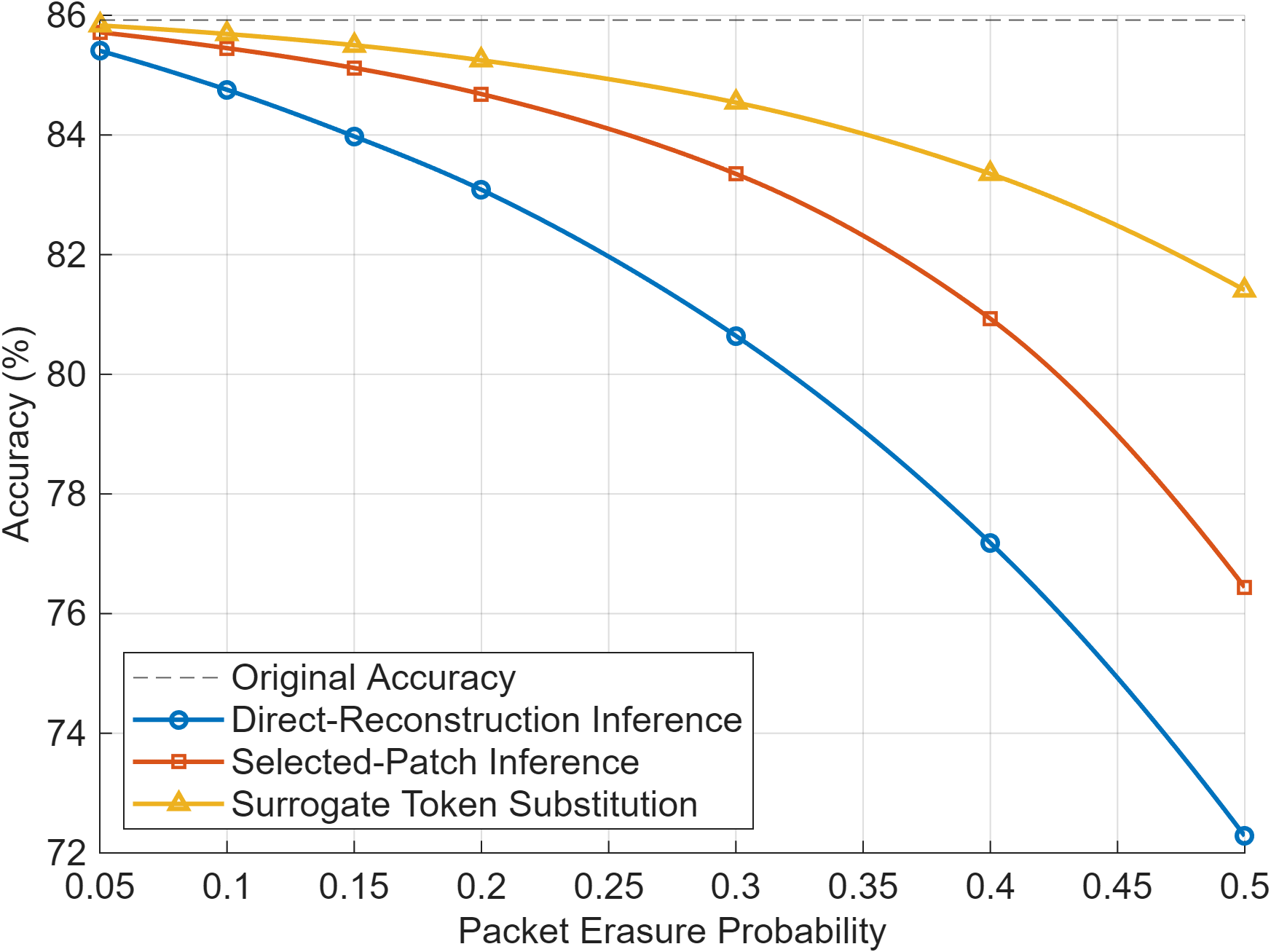}
        \caption{Classification accuracy over a packet-erasure channel. \emph{Original accuracy} corresponds to the errorless downstream performance of the proposed framework, i.e., \SI{85.92}{\%}, where $p_e = 0$. The reported accuracy is averaged over 20 independent trials.
        }
        \label{fig:exp-main-erasure}
    \end{figure}

    We evaluate the classification accuracy of DeiT-III-Large on images compressed by the LIC model with $\lambda = 0.3$ and transmitted under erasure probabilities $p_e \in \{0.05, 0.1, 0.15, 0.2, 0.3, 0.4, 0.5\}$.
    The latents in erased packets are replaced by their predicted means, and the corresponding tokens are processed by one of three server-side inference strategies: direct-reconstruction inference, selected-patch inference, and surrogate token substitution.
    To isolate the effect of erasures, selective latent transmission is not applied, fixing the bit rate at \SI{2.12}{bpp}.
    For reference, an image-level coding without packetization yields \SI{2.11}{bpp}, indicating that packetization introduces a negligible overhead.
    As shown in Fig.~\ref{fig:exp-main-erasure}, surrogate token substitution is consistently the most robust of the three strategies across the erasure probabilities.

    \subsection{Entropy-Aware Image Transmission}\label{sec:experimental-entropy}
    
    \emph{Entropy-aware image transmission} (EIT) further reduces the communication payload by allowing the client to classify sufficiently confident inputs locally without transmitting them to the server~\cite{Im2024Attention}.
    Specifically, EIT computes the min-entropy of the client-side predictive distribution, which is low for confident predictions, and invokes server-side inference only when the min-entropy is at least a threshold $\tau$.
    
    \begin{figure}[t]
        \centering
        \includegraphics[width=0.4\textwidth]{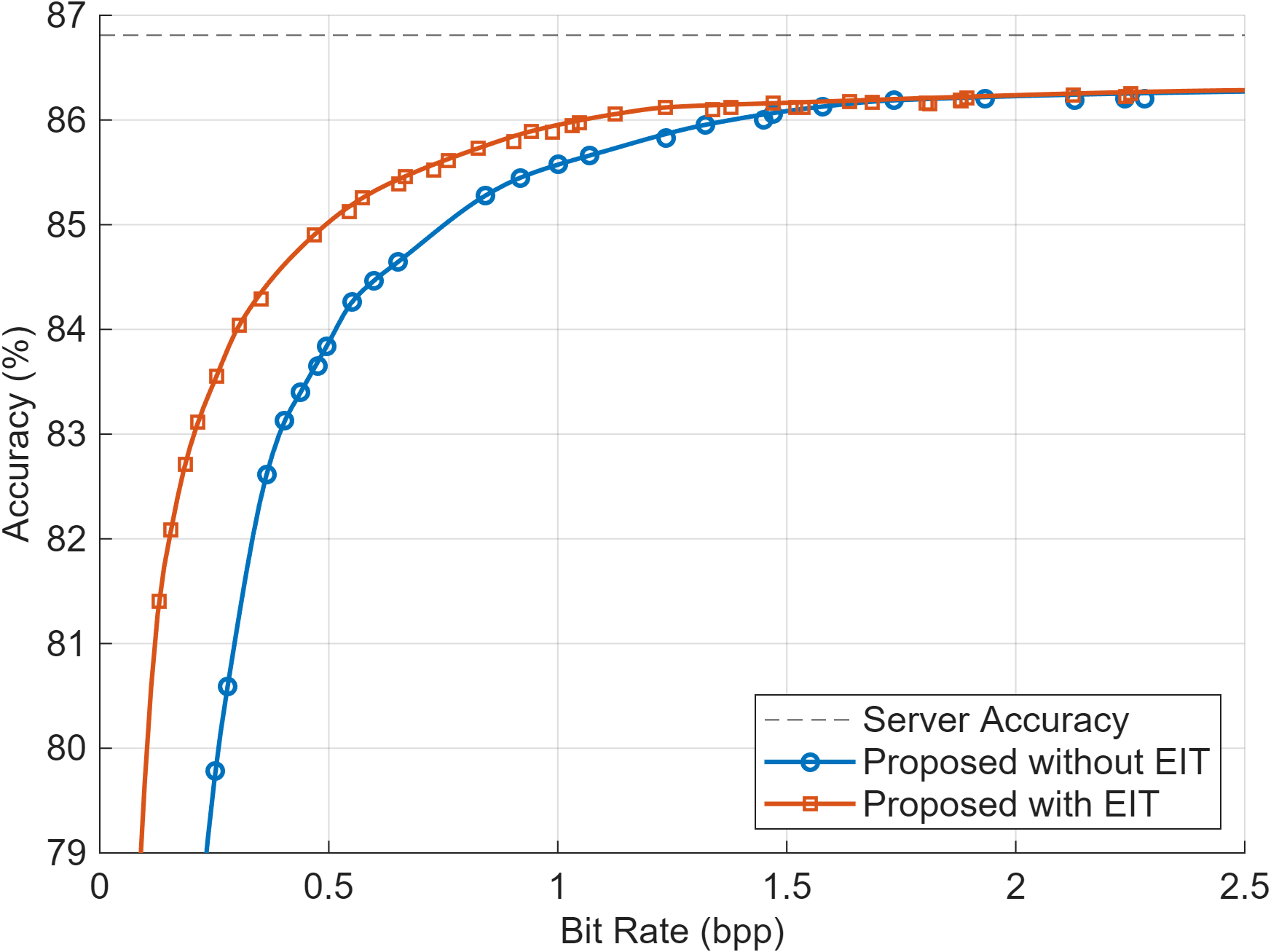}
        \caption{Trade-off between bit rate and classification accuracy of the proposed framework with and without entropy-aware image transmission (EIT), obtained over the sweep of $(\tau, \delta, \lambda)$. Both are the achievable curve formed by connecting the Pareto-optimal operating points.}
        \label{fig:exp-main-with_entropy}
    \end{figure}
    
    We sweep $\tau$ from $0.1$ to $1$, together with reduced sets of $\delta$ and $\lambda$ within the ranges of Section~\ref{sec:experimental-main}.
    As shown in Fig.~\ref{fig:exp-main-with_entropy}, EIT shifts the achievable curve toward lower bit rates: the framework with EIT attains \SI{85.89}{\%} accuracy at \SI{0.94}{bpp}, whereas approximately \SI{1.24}{bpp} is required without EIT to attain comparable accuracy.
    Since the additional computation amounts to a single classification head, EIT provides a practical complement to the proposed framework, although it is not claimed as a contribution of this work.

    \subsection{Token-Aligned Learned Image Compression}\label{sec:experimental-lic}

    \begin{figure}[t] 
        \centering
        \includegraphics[width=0.4\textwidth]{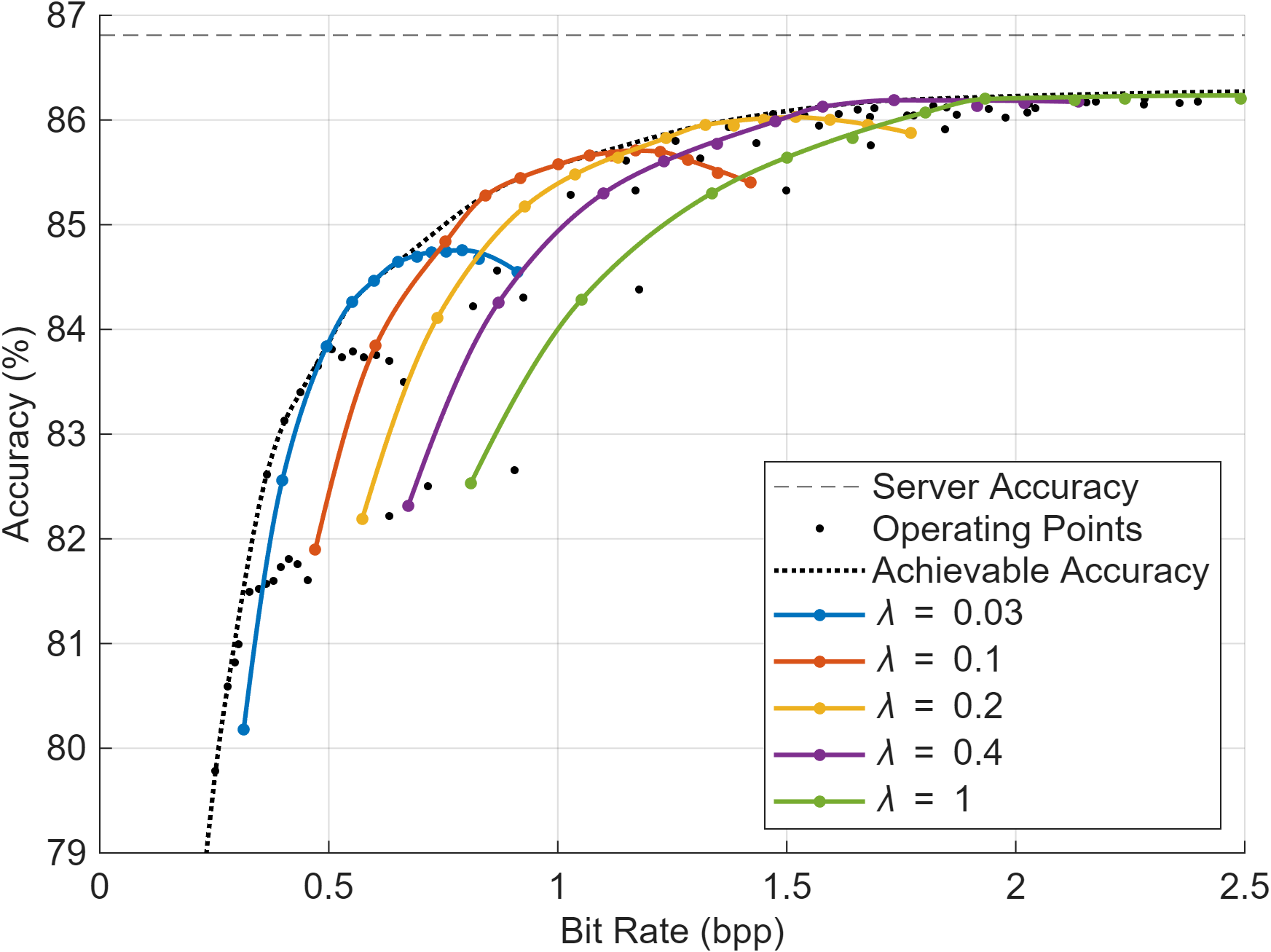}
        \caption{Achieved rate--accuracy points of the proposed framework. \emph{Achievable accuracy} is formed by connecting the Pareto-optimal operating points. Each solid line connects the operating points obtained with the same LIC model parameters while varying only $\delta$.}
        \label{fig:exp-LIC-lambdas}
    \end{figure}

    We first analyze the rate-control behavior of the proposed framework through its two hyperparameters $\delta$ and $\lambda$.
    Fig.~\ref{fig:exp-LIC-lambdas} shows the operating points obtained by varying $\delta$ and switching among LIC models trained with different $\lambda$, whose connected achievable curve enables adaptive rate control under varying channel or latency conditions.
    Adjusting only $\delta$ at inference provides fine-grained control within a narrow bit-rate range without switching the LIC model, while five LIC models with $\lambda \in \{0.03, 0.1, 0.2, 0.4, 1\}$ suffice to cover the range from \SI{0.5}{bpp} to \SI{2.5}{bpp} at near-optimal accuracy, keeping the memory cost of rate control low.

    \begin{figure}[t] 
        \centering
        \includegraphics[width=0.4\textwidth]{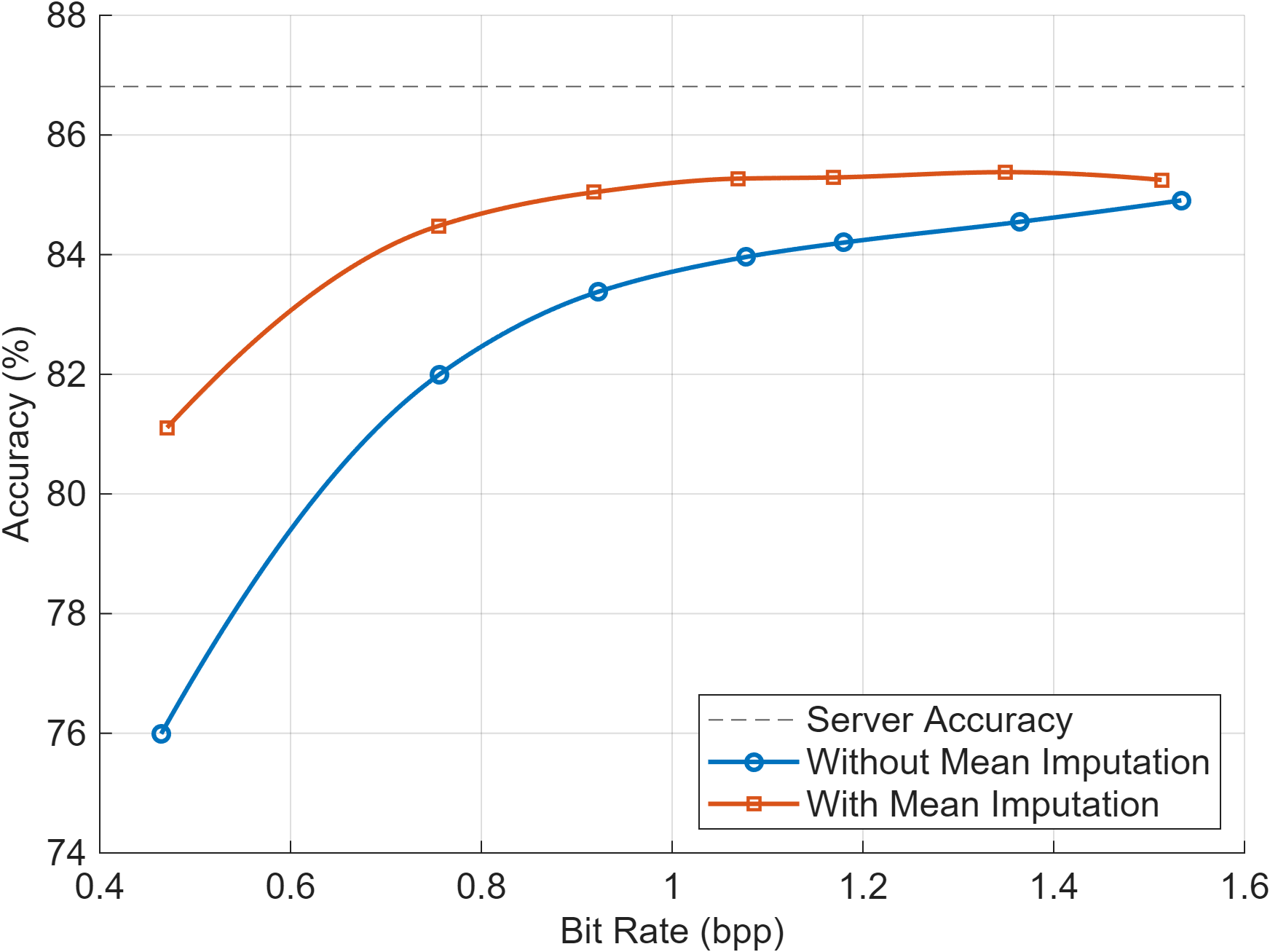}
        \caption{Trade-off between bit rate and classification accuracy for LIC models without and with mean imputation. \emph{Without mean imputation} refers to the hyperprior-based LIC model~\cite{Balle2018Variational} without mean imputation, whereas \emph{with mean imputation} denotes the model augmented with mean hyperpriors~\cite{Minnen2018Joint}, where the missing latent vectors are imputed using the predicted means. Both models are trained with $\lambda=0.1$.
        }
        \label{fig:exp-LIC-meanscale}
    \end{figure}

    To assess the effect of mean imputation, we compare hyperprior-based LIC models without and with mean imputation~\cite{Balle2018Variational,Minnen2018Joint}, using the same settings as the main experiments except that the surrogate token is trained in an LIC-agnostic manner for fairness.
    As shown in Fig.~\ref{fig:exp-LIC-meanscale}, the model with mean hyperpriors attains higher downstream accuracy.
    This gain does not come from the unsent regions, which are uniformly replaced by a surrogate token before inference, but from improved reconstruction of the selected regions: since LIC latents can influence neighboring patches, the selected patches are affected by adjacent unsent latents, and mean imputation alleviates this effect.

    \subsection{Layer-Selective Attention Rollout}\label{sec:experimental-attention}

    \begin{figure}[t] 
        \centering
        \includegraphics[width=0.4\textwidth]{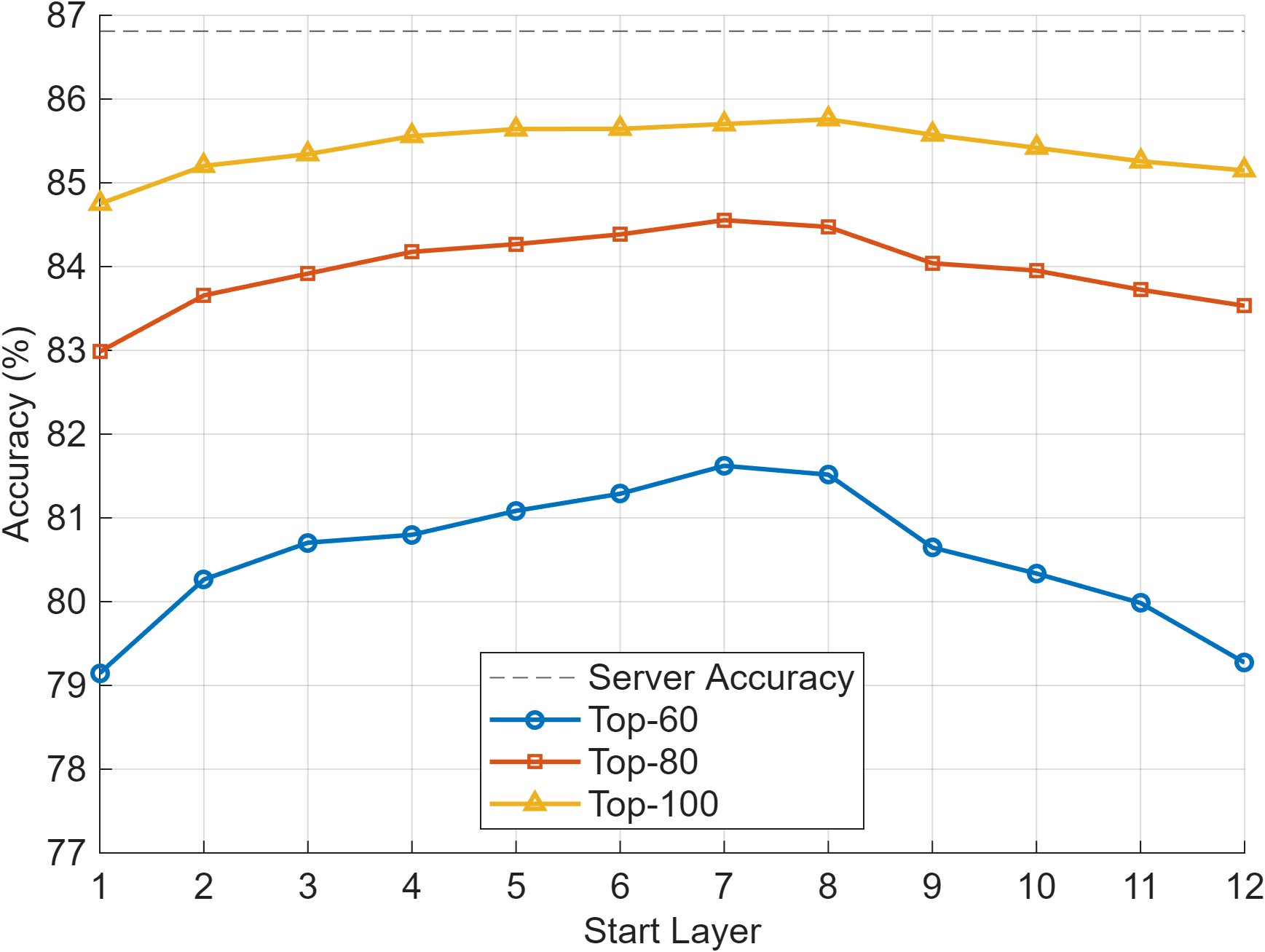}
        \caption{Selected-patch inference accuracy under layer-selective attention rollout computed with different layer ranges $(L_s, L_e)$. The end layer is fixed as $L_e = L$, where $L = 12$ for DeiT-Tiny.}
        \label{fig:exp-attention-layers}
    \end{figure}

    We first identify the most task-relevant layer range $(L_s, L_e)$, motivated by the noisy early-layer attentions observed in Fig.~\ref{fig:vis-attention}.
    Fixing $L_e = L$ and ablating the LIC module, we measure selected-patch inference accuracy in which DeiT-Tiny selects the top-$60$, top-$80$, or top-$100$ tokens by layer-selective attention rollout and DeiT-III-Large classifies from the selected patches.
    As shown in Fig.~\ref{fig:exp-attention-layers}, aggregating over the latter half of the layers, particularly with $L_s = 7$, yields the best performance in our main operating region.

    \begin{figure}[t] 
        \centering
        \includegraphics[width=0.4\textwidth]{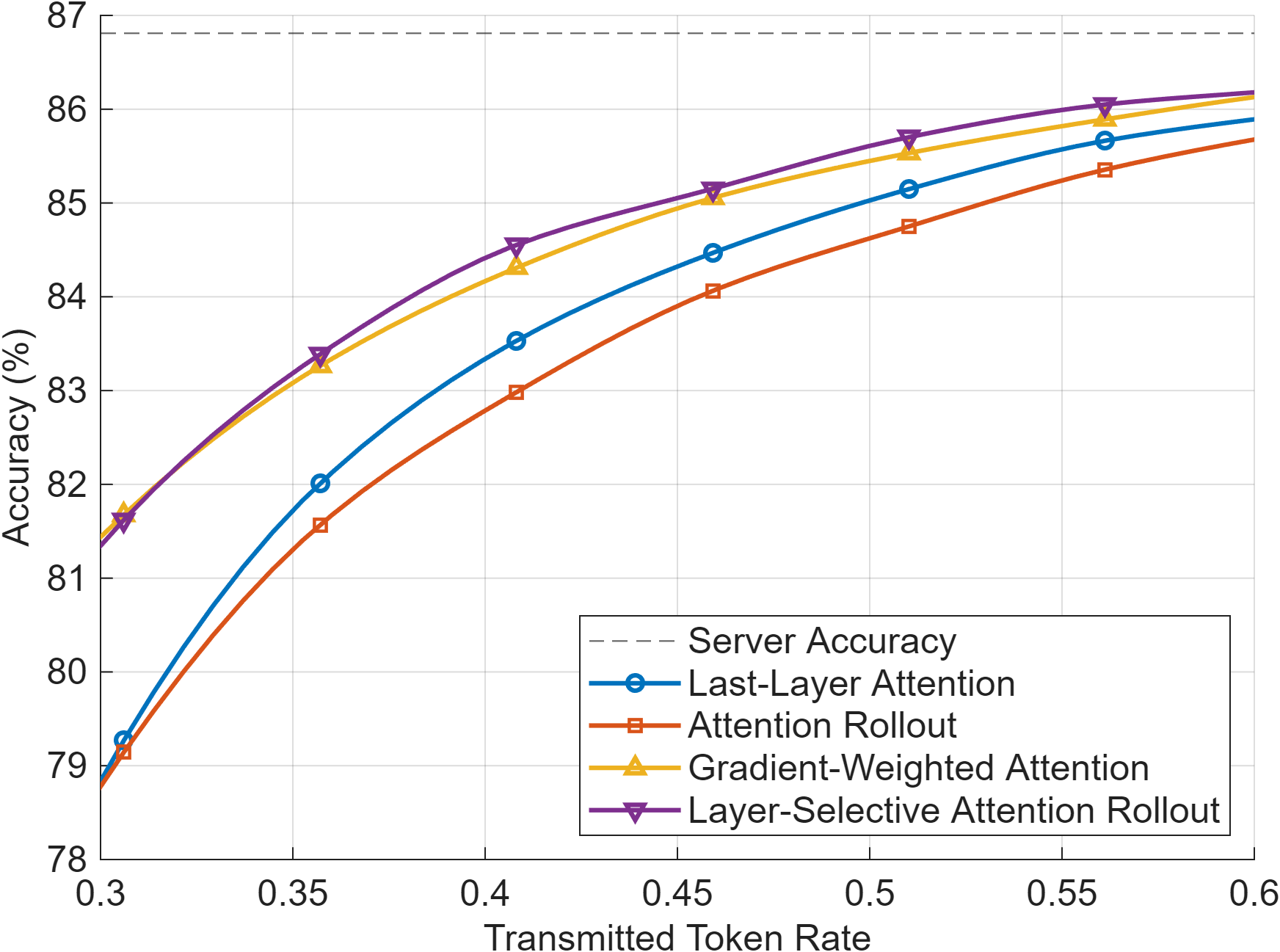}
        \caption{Selected-patch inference accuracy using different attention-guided importance. \emph{Selected token rate} is defined as the ratio of selected tokens to the total number of tokens.}
        \label{fig:exp-attention}
    \end{figure}

    We then compare layer-selective attention rollout with $(L_s, L_e) = (7, 12)$ against other attention-guided importance methods under the same setting: last-layer attention~\cite{Im2024Attention}, attention rollout~\cite{Abnar2020quantifying}, and gradient-weighted attention~\cite{Chefer2021Generic}.
    As shown in Fig.~\ref{fig:exp-attention}, layer-selective attention rollout attains the most favorable rate--accuracy trade-off in our main operating region, even without the additional gradient computation required by gradient-weighted attention.

    \begin{figure}[t]
        \centering
        \subfloat[DeiT-Small for client]{%
            \includegraphics[width=0.4\textwidth]{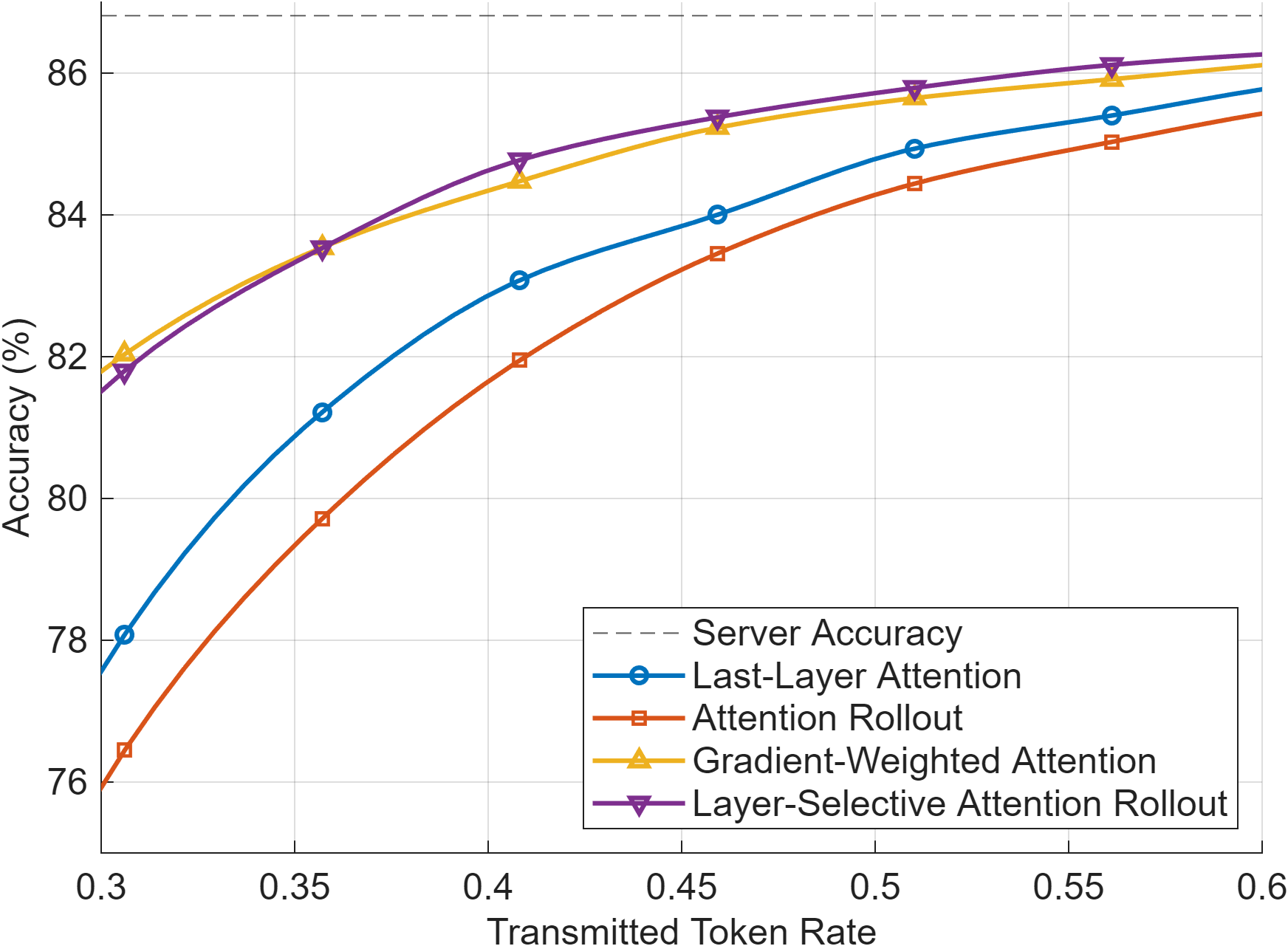}%
            \label{fig:exp-attention-small}}
        \hfil
        \subfloat[DeiT-Base for client]{%
            \includegraphics[width=0.4\textwidth]{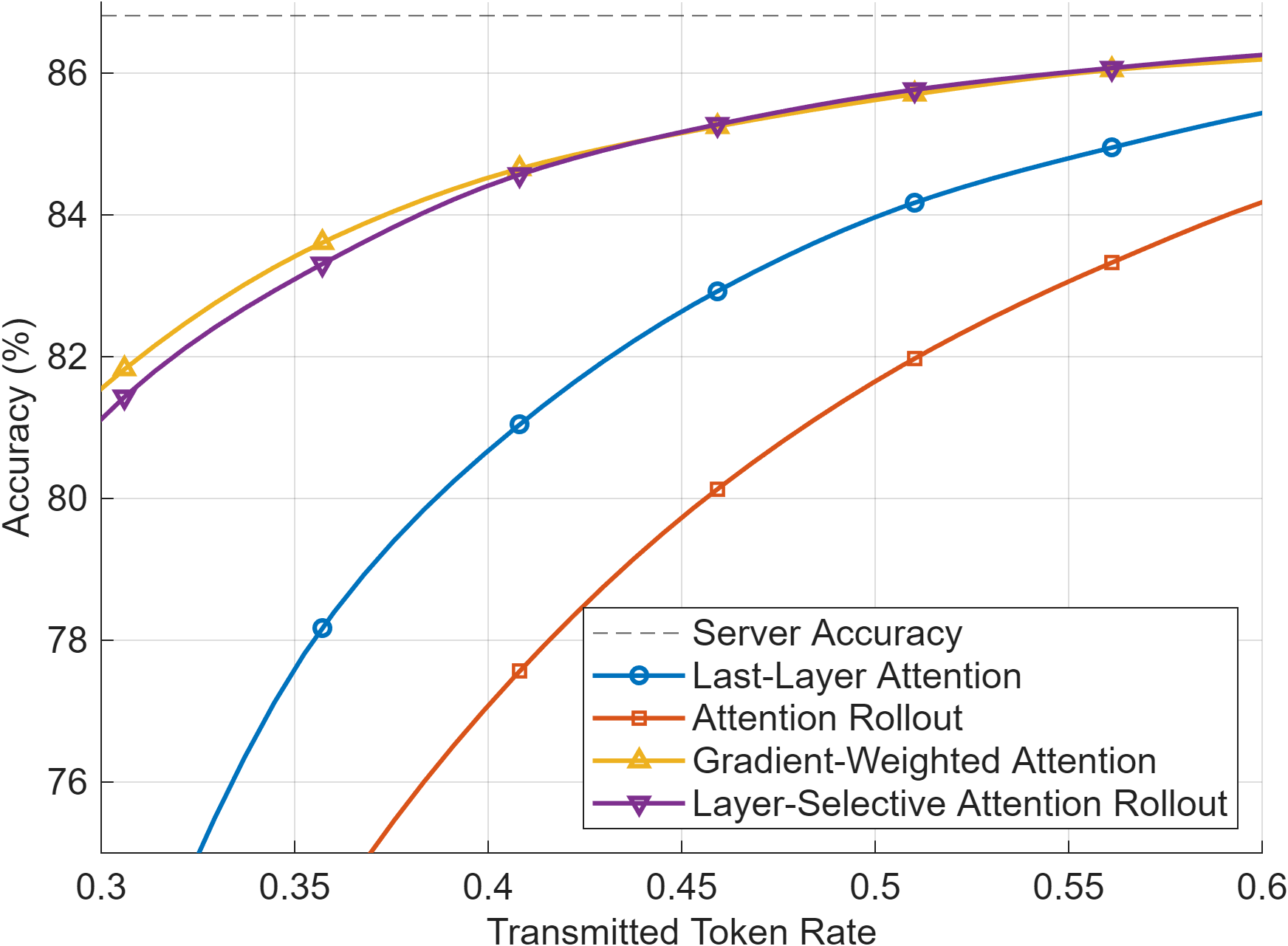}%
            \label{fig:exp-attention-base}}
        \caption{Selected-patch inference accuracy using different attention-guided importance with the client-side ViT replaced by (a) DeiT-Small and (b) DeiT-Base, both having $L = 12$ transformer layers. Layer-selective attention rollout is computed over $(L_s, L_e) = (5, 12)$ and $(7, 12)$, respectively.}
        \label{fig:exp-attention-model}
    \end{figure}

    Moreover, we verify that the effectiveness of layer-selective attention rollout is not specific to the choice of client-side ViT by repeating the comparison with DeiT-Tiny replaced by DeiT-Small and DeiT-Base.
    Both backbones comprise $L = 12$ transformer layers, for which we set $(L_s, L_e) = (5, 12)$ and $(7, 12)$, respectively, searched by the same procedure used for DeiT-Tiny.
    As shown in Fig.~\ref{fig:exp-attention-model}, layer-selective attention rollout consistently outperforms last-layer attention and attention rollout, and performs comparably to gradient-weighted attention on both backbones.
    Crucially, layer-selective attention rollout attains this accuracy with only a single forward pass, whereas the gradient computation required by gradient-weighted attention becomes increasingly impractical for these larger backbones---making layer-selective attention rollout the more practical choice at comparable accuracy.

    \subsection{Surrogate Token Substitution}\label{sec:experimental-surrogate}
    
    The surrogate token is trained by emulating the proposed pipeline under training hyperparameters $\delta_{\mathrm{sur}}$ and $\lambda_{\mathrm{sur}}$, which set the token-selection threshold and the LIC operating point during training, respectively; the detailed settings are given in Section~\ref{sec:experimental-settings}.
    Because these hyperparameters determine the quality of training-time inputs, they can affect both training stability and final accuracy.

    \begin{figure}[t] 
        \centering
        \includegraphics[width=0.4\textwidth]{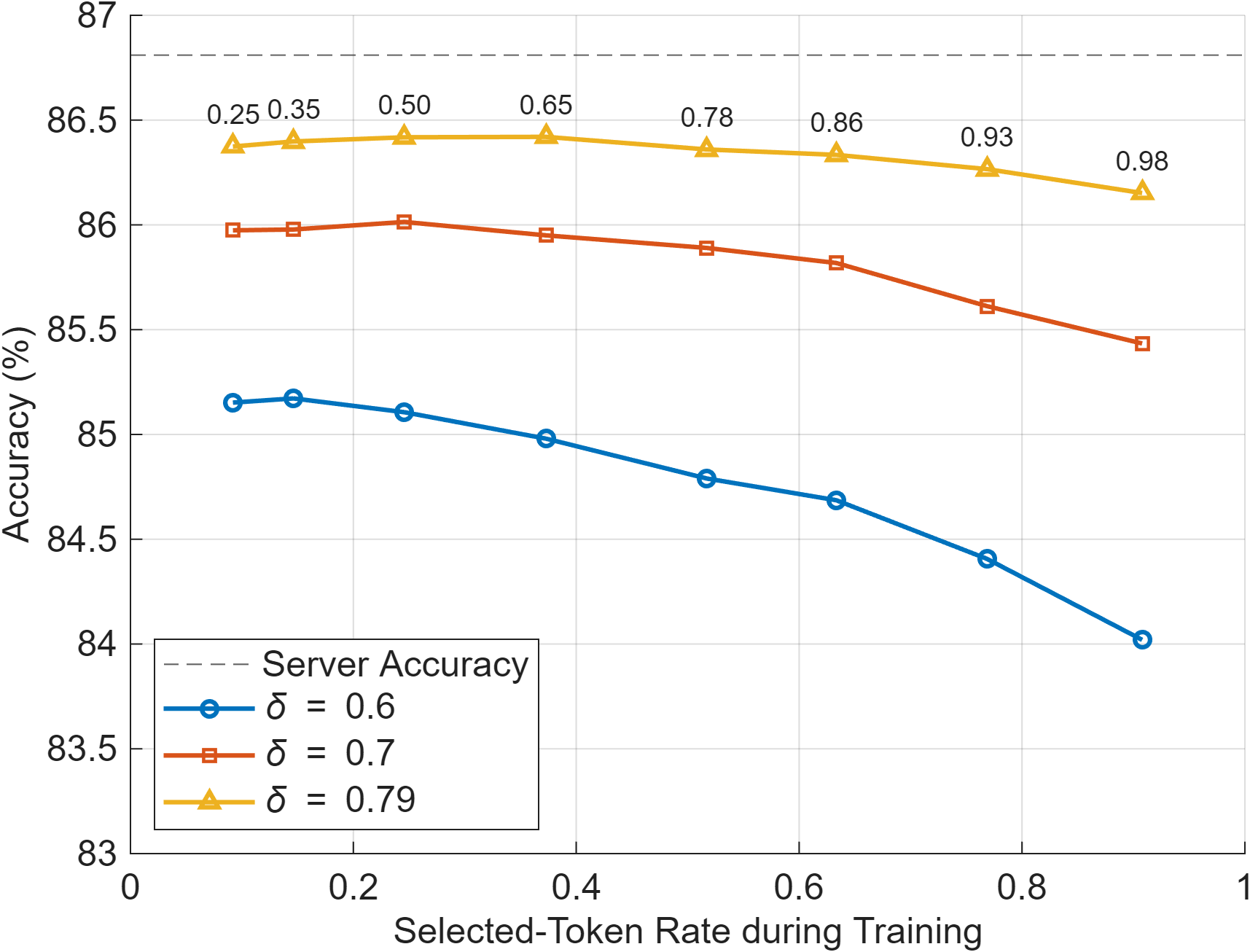}
        \caption{Evaluation accuracy for surrogate tokens trained under different token-selection thresholds $\delta_{\mathrm{sur}}$. We evaluate under the test-time thresholds $\delta \in \{0.6, 0.7, 0.79\}$. The annotations indicate the corresponding values of $\delta_{\mathrm{sur}}$ for each training token rate.}
        \label{fig:exp-surrogate-training_tokens}
    \end{figure}

    We first determine $\delta_{\mathrm{sur}}$ by ablating the LIC module and training a separate surrogate token for each value of $\delta_{\mathrm{sur}}$ from $0.25$ to $0.98$.
    Fig.~\ref{fig:exp-surrogate-training_tokens} evaluates these tokens at test-time thresholds $\delta \in \{0.6, 0.7, 0.79\}$ in our main operating region, showing that $\delta_{\mathrm{sur}} = 0.35$ consistently attains near-optimal accuracy despite the mismatch between $\delta$ and $\delta_{\mathrm{sur}}$.

    \begin{figure}[t]
        \centering
        \subfloat[Server-side inference strategies]{%
            \includegraphics[width=0.4\textwidth]{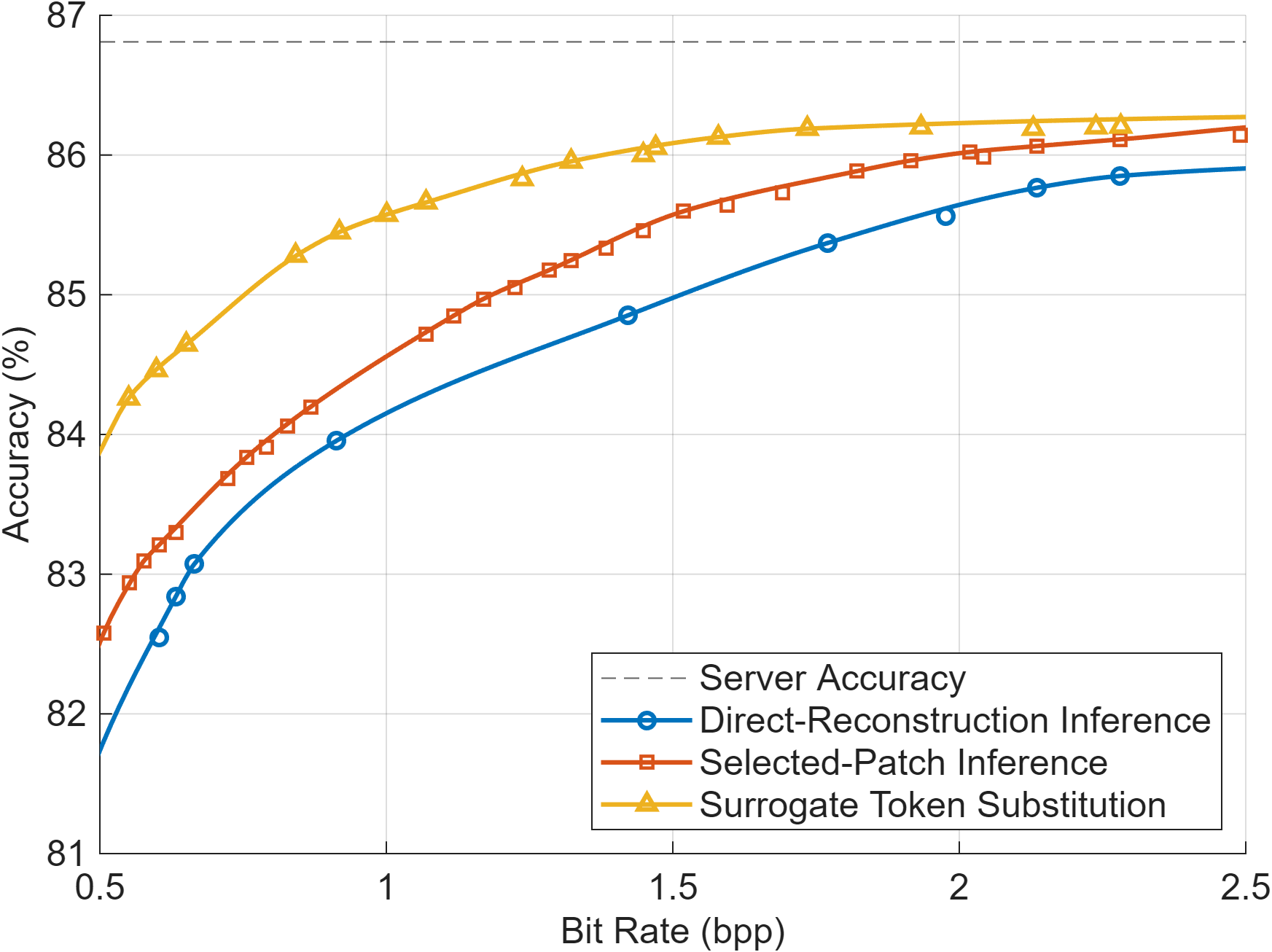}%
            \label{fig:exp-surrogate}}
        \hfil
        \subfloat[LIC-aware vs.\ LIC-agnostic training]{%
            \includegraphics[width=0.4\textwidth]{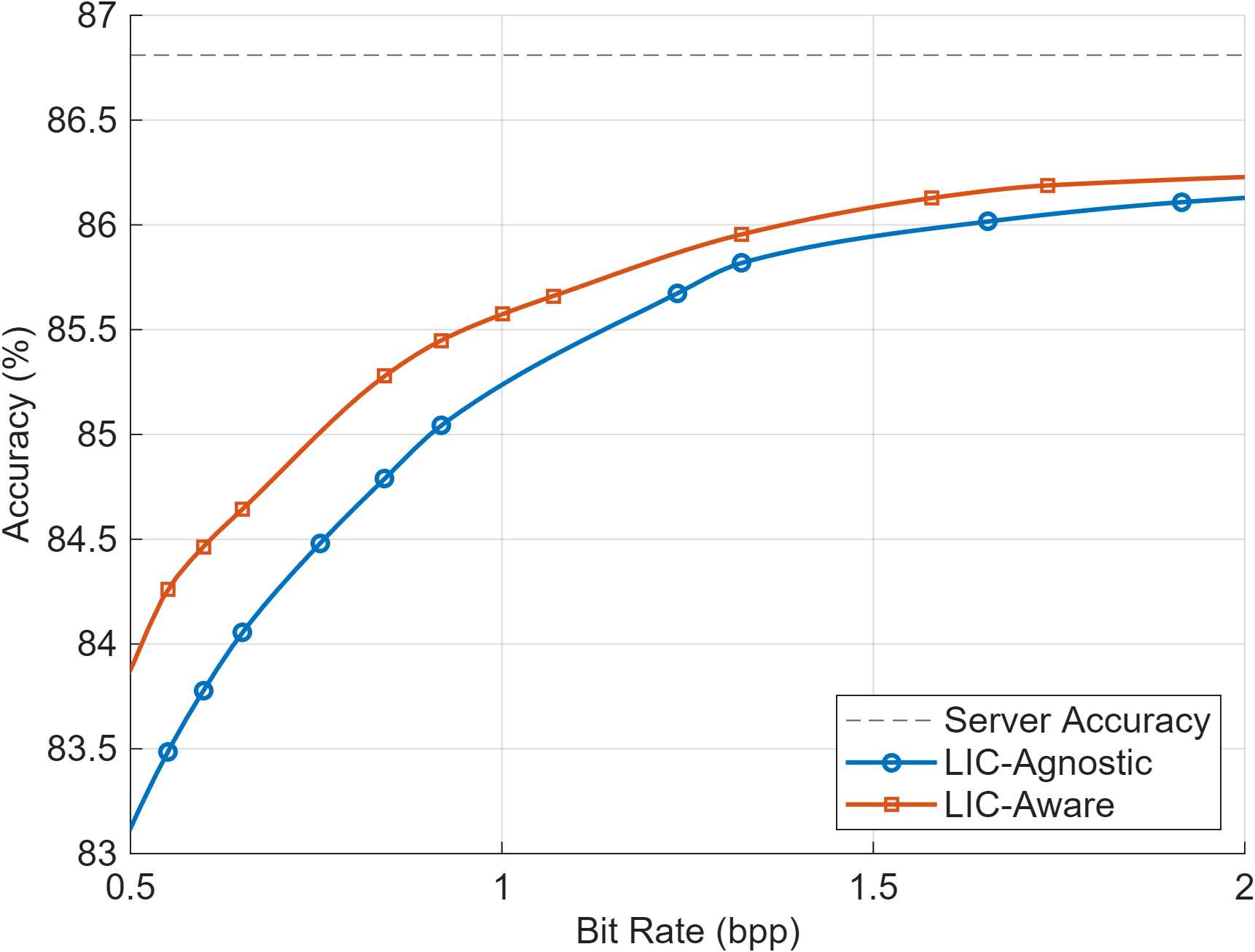}%
            \label{fig:exp-surrogate-LIC_awareness}}
        \caption{Trade-off between bit rate and classification accuracy of surrogate token substitution. (a) Comparison of server-side inference strategies, where the surrogate token is trained with $(\delta_{\mathrm{sur}}, \lambda_{\mathrm{sur}}) = (0.35, 0.03)$. (b) Comparison of surrogate tokens trained with and without the LIC module.}
        \label{fig:exp-surrogate-subplot}
    \end{figure}

    Fixing $\delta_{\mathrm{sur}} = 0.35$, we determine $\lambda_{\mathrm{sur}} = 0.03$ in the same manner.
    With these hyperparameters, Fig.~\ref{fig:exp-surrogate-subplot}\subref{fig:exp-surrogate} shows that surrogate token substitution outperforms both adaptation-free baselines.
    Furthermore, Fig.~~\ref{fig:exp-surrogate-subplot}\subref{fig:exp-surrogate-LIC_awareness} shows that the \emph{LIC-aware} surrogate token, trained with $(\delta_{\mathrm{sur}}, \lambda_{\mathrm{sur}}) = (0.35, 0.03)$, consistently outperforms the \emph{LIC-agnostic} counterpart trained with $\delta_{\mathrm{sur}} = 0.35$ alone, indicating that incorporating LIC awareness during training yields additional gains.

    \section{Conclusion}\label{sec:conclusion}

    We proposed a modular, token-oriented semantic communication framework that coordinates pretrained client-side ViT, LIC, and server-side ViT models for client--server collaborative inference.
    To improve the rate--accuracy trade-off, we developed three technical components: 
    (1) \emph{token-aligned learned image compression}, which uses token-level task relevance to selectively transmit spatially aligned LIC latents; 
    (2) \emph{layer-selective attention rollout}, which efficiently estimates token relevance from a selected range of attention layers; and 
    (3) \emph{learnable surrogate token substitution}, which compensates for missing visual information at the server. 
    Experimental results demonstrated that the proposed framework achieves a more favorable rate--accuracy trade-off than recent token-oriented semantic communication schemes, hand-crafted codecs, and learned image compression models. 
    Overall, these results indicate that token-aligned compression of the input image, rather than direct compression of token representations, is a promising direction for modular token-level semantic communication in resource-constrained edge AI systems.
    
    \appendices
	
    \bibliographystyle{IEEEtran}
    \bibliography{abrv,mybib}

\end{document}